\documentclass{jpp}
\usepackage{graphicx}

\usepackage[utf8]{inputenc}
\usepackage[T1]{fontenc}
\usepackage{amsmath}

\usepackage{amsfonts}
\usepackage[caption=false,justification=raggedright]{subfig}
\usepackage[export]{adjustbox}
\usepackage{xcolor}
\usepackage{amssymb}
\usepackage{physics}
\usepackage{xcolor}
\usepackage{hyperref}
\usepackage[caption=false,justification=raggedright]{subfig}
\usepackage{stmaryrd}

\usepackage{tensor}
\usepackage{wrapfig}
\newcommand{\gke}{\texttt{Gkeyll}}
\newcommand{\eqr}[1]{Eq.\thinspace(#1)}
\newcommand{\pfrac}[2]{\frac{\partial #1}{\partial #2}}

\newcommand{\pfraca}[1]{\frac{\partial}{\partial #1}}

\newcommand{\mvec}[1]{\mathbf{#1}}

\newcommand{\gcs}{\nabla_{\mvec{x}}}

\newcommand{\basis}[1]{\mvec{e}_{#1}}

\newcommand{\dbasis}[1]{\mvec{e}^{#1}}

\newcommand{\ndbasis}[1]{\hat{\mvec{e}}^{#1}}

\newcommand{\gbasis}[2]{\mvec{#1}_{#2}}
\newcommand{\gdbasis}[2]{\mvec{#1}^{#2}}

\newcommand{\incfig}{\centering\includegraphics}
\setkeys{Gin}{width=0.9\linewidth,keepaspectratio}

\newcommand{\buni}{\mvec{b}}

\shorttitle{Field Aligned Coordinate Systems for Gyrokinetic Simulations}
\shortauthor{A. Shukla, A. Hakim, J. Juno, G. W. Hammett, and M. Francisquez}

\title{Constructing Field Aligned Coordinate Systems for Gyrokinetic
  Simulations of Tokamaks in X-point Geometries}

\author{Akash Shukla\aff{1}
  \corresp{\email{akashukla@utexas.edu}},
  A. Hakim\aff{2},
  J. Juno\aff{2},
  G. W. Hammett\aff{2},
  \and M.  Francisquez\aff{2}
 }

\affiliation{%
  \aff{1}Institute for Fusion Studies, University of Texas at Austin,
  2515 Speedway, Austin, TX 78712, USA%
  \aff{2}Princeton Plasma Physics Laboratory, 100 Stellarator Rd, Princeton, NJ
  08540, USA}

\begin{document}

\maketitle

\begin{abstract}
    Structures in tokamak plasmas are elongated along the direction of the magnetic field and short in the directions perpendicular to the magnetic field. Many tokamak simulation codes take advantage of this by using a field aligned coordinate system. However, field aligned coordinate systems have a coordinate singularity at magnetic X-points where the poloidal magnetic field vanishes, which makes it difficult to use field aligned coordinate systems when simulating the core and scrape-off layer (SOL) simultaneously. Here we present an algorithm for computing geometric quantities in a standard field aligned coordinate system that avoids the singularity and allows one to conduct 2D axisymmetric simulations in X-point geometries. We demonstrate the efficacy of this algorithm with an example simulation of the Spherical Tokamak for Energy Production (STEP).
\end{abstract}

\section{Introduction}

    Structures in tokamak plasmas are anisotropic: they are elongated along the field line but short perpendicular to it. Many tokamak simulation codes, especially core codes such as GS2~\citep{Dorland2000,GS2-zenodo}, GENE~\citep{Jenko2000,Gorler2011} and GYRO~\citep{Candy2010, Candy2009}, take advantage of this by using a field aligned coordinate system; the field aligned coordinate system allows for coarse resolution along the field line reducing computational expense~\citep{beer95}. However, field aligned coordinate systems have a coordinate singularity at magnetic X-points where the poloidal magnetic field vanishes, so using field-aligned coordinate systems in edge codes which simulate the core and scrape-off layer (SOL) simultaneously is more difficult~\citep{STEGMEIR2016139, leddy2017}. 

    There have been a variety of approaches to handling the coordinate singularity at the X-point. BOUT++ handles it by using multiple blocks, each with a field aligned coordinate system, 
    and avoiding the calculation of geometric quantities at the X-point. 
    COGENT uses multiple blocks each with a coordinate system that is flux aligned except near the X-point where they overlap. A high order interpolation scheme is used to transfer information between the overlapping regions of each block~\citep{MCCORQUODALE2015181, Dorf16}.

    Other edge gyrokinetic codes such as GENE-X~\citep{Michels21} have abandoned field and flux aligned coordinates in favor of the Flux-Coordinate-Independent (FCI) approach because of the difficulty of dealing with the singularity at the X-point. The FCI approach breaks the simulation domain into a series of poloidal planes which do not have a field aligned coordinate system and employs a field-line following discretization of the parallel derivative operator to minimize the number of poloidal planes needed. Interpolation within the poloidal plane is required to compute the parallel derivatives~\citep{HARIRI20132419, STEGMEIR2016139, Stegmeir_2018}.
    
    Here we present an algorithm for computing geometric quantities in a standard field aligned coordinate system that avoids the singularity at the X-point. We employ a multi-block approach where each block conforms to the separatrix leaving no gap around the X-point. Our numerical scheme allows us to avoid calculation of any geometric quantities or fluxes at the X-point while still having block corners at the X-point. We implement and test this algorithm in the gyrokinetic model in the \gke\ simulation framework\citep{Mana25, Shukla25, Mandell2020, Ammar2019}.

    The rest of the paper is organized as follows: in section~\ref{sec:clebsch} we give background on the Clebsch representation of magnetic fields and field aligned coordinates,
    in section~\ref{sec:transforms} we present the equations of our gyrokinetic model in a field aligned coordinate system, 
    in section~\ref{sec:coordinates} we detail the coordinate system we employ, 
    in section~\ref{sec:discretization} we show how the spatial discretization of our algorithm avoids the cooordinate singularity at the X-point,
    and in section~\ref{sec:computation} we describe how we generate simulation grids and calculate geometric quantities and also show example grids. 
    Finally, in section~\ref{sec:example}, we show an example 2D axisymmetric gyrokinetic simulation of STEP~\citep{Karhunen2024} using this method.
    In Appendix~\ref{sec:appendix} we describe geometric consistency requirements of numerical schemes to solve the advection equation, including for the multi-block case. The same consistency conditions are also required for the gyrokinetic equations which is an advection equation in phase-space.

\section{Coordinate Systems for Magnetized Plasma Simulations}
\label{sec:clebsch}

As is well known, in certain situations (described below), we can
write the magnetic field in the Clebsch representation~\citep{DHaessler}
\begin{align}
  \mvec{B} = C(\mvec{x})\nabla\psi\times\nabla\alpha \label{eq:clebsch}
\end{align}
where, $C(\mvec{x})$ $\psi(\mvec{x})$ and $\alpha(\mvec{x})$ are
scalar functions of the position vector $\mvec{x}$. The divergence
constraint $\nabla\cdot\mvec{B} = 0$ requires
\begin{align}
  \nabla\cdot\mvec{B} = \frac{1}{C} \mvec{B}\cdot\nabla C = 0.
\end{align}
However, not all magnetic field configurations can be described by the
Clebsch representation: the field-lines of Clebsch-representable
magnetic fields are \emph{integrable} and hence enforce some stringent
constraints on the type of fields that can be described in this
way\footnote{A generalized Clebsch representation of the form
  $\mvec{B} = \nabla\psi_1\times\nabla\alpha_1 +
  \nabla\psi_2\times\nabla\alpha_2$ allows representing arbitrary
  magnetic fields, including ones in which the field-lines are not
  integrable. However, these are not useful to construct coordinate
  systems.}. Thankfully, for tokamaks, where the fields are
axisymmetric, or in regions of stellarators with nested flux surfaces,
such representations can be found. Hence, in this paper we will
restrict ourselves to such magnetic configurations.

The importance of the Clebsch representation (when it exists) is that
the two vectors $\nabla\psi$ and $\nabla\alpha$ can be used as the
\emph{dual basis vectors} (contravariant basis) of a field-line
following coordinate system. To understand what this means and
establish notation for rest of the paper consider an arbitrary
coordinate transform given by the \emph{invertible} map
\begin{align}
  \mvec{x} = \mvec{x}(z^1, z^2, z^3)
\end{align}
where $(z^1, z^2, z^3)$ are computational coordinates. This maps a
rectangular region in $\mathbb{R}^3$ to a (generally non-rectangular)
region of physical space. Once this mapping is known then we can compute the
\emph{tangent vectors}
\begin{align}
  \basis{i} = \pfrac{\mvec{x}}{z^i}
\end{align}
and the \emph{dual vectors} $\dbasis{i}$ defined implicitly by the
relation
\begin{align}
  \dbasis{i}\cdot\basis{j} = \delta\indices{^i_j}.
\end{align}
If the inverse mapping $z^i = z^i(\mvec{x})$ is known, then we can
show that $\dbasis{i} = \nabla z^i$. At each point $\mvec{x}$ either
the tangents or duals form a linearly independent set of vectors and
hence can be used to represent vector and tensor quantities at that
point. For example, a vector $\mvec{a}$ can be written as
\begin{align}
  \mvec{a} = a^i \basis{i} = a_i \dbasis{i}
\end{align}
where $a^i = \mvec{a}\cdot\dbasis{i}$, $a_i = \mvec{a}\cdot\basis{i}$
and we have assumed the summation convention over repeated
indices. Once the tangent and dual vectors are determined we can
compute the covariant and contravariant components of the metric
tensor as
\begin{align}
  g_{ij} &= \basis{i}\cdot\basis{j} \\
  g^{ij} &= \dbasis{i}\cdot\dbasis{j}.
\end{align}

Defining the \emph{Jacobian} (volume element) of the transform
$J_c = \basis{1}\cdot(\basis{2}\times\basis{3})$ we can
easily derive the explicit expressions for the duals:
\begin{align}
  \dbasis{1} &= \frac{1}{J_c} \basis{2}\times\basis{3} \\
  \dbasis{2} &= \frac{1}{J_c} \basis{3}\times\basis{1} \\
  \dbasis{3} &= \frac{1}{J_c} \basis{1}\times\basis{2}.
\end{align}
From this we also see that
$J_c^{-1} = \dbasis{1}\cdot(\dbasis{2}\times\dbasis{3}) =
\nabla z^1\cdot(\nabla z^2\times\nabla z^3)$. We assume that the basis
are arranged such that $J_c > 0$. 

As we need the mapping to be invertible we must ensure that
$J_c(\mvec{x})$ does not vanish anywhere in the domain. At
the $X$- and $O$-points of a tokamak configurations, however, we have
$J_c = 0$ for field-line following coordinates, that is, the
coordinate system is non-invertible. We get around this issue by
ensuring that we \emph{do not} compute any geometrical quantities or
numerical fluxes at these isolated singular points in the domain. The
use of a high-order scheme (we use the discontinuous Galerkin scheme)
that uses interior (to surfaces and volumes) quadrature nodes where
numerical fluxes are computed, automatically ensures this, allowing us
to work with coordinate systems that have singularities at a finite
set of isolated points. However, despite not computing any geometric or physical quantity at the $X$- or $O$-points, we ensure a corner node lies exactly there, producing an accurate representation of the geometry, without any ``holes''.

Identifying the dual vectors as $\dbasis{i} = \nabla z^i$, we can see why the Clebsch
form \eqr{\ref{eq:clebsch}} is useful: once we find the Clebsch form
we can \emph{construct} a coordinate system (as described later in
this paper) such that the resulting mapping has
$\dbasis{1} = \nabla\psi$ and $\dbasis{2} = \nabla\alpha$. With this,
the two scalar function $z^1 = \psi$ and $z^2 = \alpha$ would be two
of the three computational coordinates. The choice of the third
coordinate, $z^3 = \theta$, called the field-line coordinate, can then
be made independently.

Now, as
$\mvec{B} = C \nabla\psi\times\nabla\alpha = C
\dbasis{1}\times\dbasis{2}$ we must have
\begin{align}
  \mvec{B}\cdot\dbasis{1} = \mvec{B}\cdot\dbasis{2} = 0
\end{align}
and hence
\begin{align}
  \mvec{B} = (\mvec{B}\cdot\dbasis{3})\basis{3} = \frac{C}{J_c} \basis{3}.
\end{align}
From this we get a relation between the Jacobian, the magnitude of the
magnetic field and the $g_{33}$:
\begin{align}
  J_c B = C \sqrt{g_{33}}.
\end{align}

In these \emph{field-line following} coordinates the magnetic field
always points in the direction of $\basis{3}$. The unit vector in the
direction of the magnetic field is
\begin{align}
  \buni = \frac{\basis{3}}{\| \basis{3} \|}
\end{align}

The choice of these field-line following coordinates, is \emph{not},
in general, global or unique, and depends on the topologically
distinct regions that need to be included in a simulation. 
In general, a single mapping is not usually enough to cover all of the physical region of interest, and hence 
several maps are needed that between them cover the physical domain. 
For simple
devices, like the magnetic mirror, a single coordinate map is enough
to grid the complete domain. However, for tokamaks we usually have to
divide the physical domain into multiple regions, at least one for
each topologically distinct region, and construct field-line following
coordinates specific to each region. For example, for a double-null
configuration we have to construct separate coordinate systems in the
outer and inner scrape-off-layers (SOLs), the upper and lower private
flux (PF) regions and the core region. In our implementation, in fact,
for double-null configurations, we generate 
five maps to ensure a reasonably smooth grid that includes the core, the
SOLs and the private-flux regions. We refer to the assembly of grids that covers the entire physical region of interest as a multi-block grid.

\section{Transforms of the Gyrokinetic Equation}
\label{sec:transforms}
\subsection{The Gyrokinetic Equations}

The electrostatic gyrokinetic equation can be written as a Hamiltonian
system
\begin{align}
  \pfrac{f}{t} + \{f,H\} = 0 \label{eq:gen-ham}
\end{align}
where $f$ is the distribution function and $H$ is the Hamiltonian. In conservative form we can write this
as
\begin{align}
  \pfrac{(\mathcal{J}f)}{t} + 
  \gcs\cdot(\mathcal{J}\dot{\mvec{x}} f)
  +
  \pfraca{v_\parallel}
  (\mathcal{J}\dot{v}_\parallel f) = 0
  \label{eq:gen-ham-cons}
\end{align}
where $v_\parallel$ is the velocity parallel to the magnetic field, $\mu$ is the magnetic moment, $\dot{\mvec{x}} = \{\mvec{x},H\}$, $\dot{v}_\parallel = \{v_\parallel,H\}$ and
$\mathcal{J} = B_\parallel^*/m$. Further, for any two phase-space functions
$f(\mvec{x},v_\parallel,\mu)$ and $g(\mvec{x},v_\parallel,\mu)$ the
Poisson bracket given by
\begin{align}
  \{f,g\}
  =
  \frac{\mvec{B}^*}{m B_\parallel^*}
  \cdot
  \left(
  \gcs f \pfrac{g}{v_\parallel}
  -
  \pfrac{f}{v_\parallel}
  \gcs g
  \right)
  -
  \frac{\buni}{q B_\parallel^*}
  \times
  \gcs f \cdot \gcs g
\end{align}
where $\mvec{B}^* = \mvec{B} + (m v_\parallel/q) \gcs\times\buni$ and
$B_\parallel^* = \buni\cdot\mvec{B}^* \approx B$. The Hamiltonian is
\begin{align}
  H = \frac{1}{2} m v_\parallel^2 + \mu B + q \phi,
\end{align}
where $m$ is the species' mass, $q$ is the species' charge, and $\phi$ is the electrostatic potential. Substituting the Hamiltonian into the Poisson bracket, we get get
\begin{align}
  \{f,H\}
  &=
  \frac{\mvec{B}^*}{m B_\parallel^*}
  \cdot
  \left(
  m v_\parallel
  \gcs f
  -
  \pfrac{f}{v_\parallel}
  \gcs H
  \right)
  -
  \frac{\buni}{q B_\parallel^*}
  \times
  \gcs f \cdot \gcs H \\
  &=
  \frac{\mvec{B}^*}{m B_\parallel^*}
  \cdot
  \left(
  m v_\parallel
  \gcs f
  -
  \pfrac{f}{v_\parallel}
  \gcs H
  \right)
  +
  \frac{\buni}{q B_\parallel^*}
  \times
  \gcs H \cdot \gcs f
\end{align}
where
\begin{align}
  \gcs H
  =
  \mu\gcs B + q\gcs\phi.
\end{align}
The characteristics are
\begin{align}
  \dot{\mvec{x}} = \{\mvec{x},H\} =
  \frac{\mvec{B}^*}{B_\parallel^*} v_\parallel
  +
  \frac{\buni}{q B_\parallel^*}
  \times
  \gcs H
\end{align}
and
\begin{align}
  \dot{v}_\parallel = \{{v}_\parallel,H\} =
  -\frac{\mvec{B}^*}{m B_\parallel^*}\cdot\gcs H.
\end{align}

The electrostatic potential $\phi$ is determined by the gyrokinetic Poisson equation (also sometimes called the quasinetrality equtaion)
\begin{align}
  -\gcs\cdot
  \left(
  \varepsilon_\perp \nabla_\perp \phi
  \right)
  =
  \sum_s q_s \int \mathcal{J} f_s \thinspace d^3\mvec{v}
\end{align}
where $\varepsilon_\perp(\mvec{x})$ is a polarization tensor and $d^3\mathbf v = d\mu dv_\parallel$ indicates integration of velocity space. The
operator $\nabla_\perp$ is defined as
\begin{align}
  \nabla_\perp = \gcs - \buni(\buni\cdot\gcs).
\end{align}

Until this point we have written all equations in an coordinate
independent form. Now we introduce coordinates. Consider transforming
the configuration space coordinates as
\begin{align}
  \mvec{x} = \mvec{x}(z^1,z^2,z^3)
\end{align}
where $(z^1,z^2,z^3)$ are computational coordinates. From this
mapping, as we discussed above, we can compute the tangent vectors
\begin{align}
  \basis{i} = \pfrac{\mvec{x}}{z^i}
\end{align}
and the duals from $\basis{i}\cdot\dbasis{j} = \delta^i_j$, and the
co- and contravariant components of the metric-tensor
$g_{ij} = \basis{i}\cdot\basis{j}$,
$g^{ij} = \dbasis{i}\cdot\dbasis{j}$.

One we have the tangents and duals, we can construct the fundamental
vector derivative operator
\begin{align}
  \gcs  = \dbasis{i}\frac{\partial}{\partial z^i}.
\end{align}
This operator is enough now to write the equations in aribitrary
coordinate systems. To ease the derivations we need the identities
\begin{align}
  \gcs\cdot\mvec{U}
  =
  \frac{1}{J_c}\frac{\partial}{\partial z^i}\left(J_c
  \dbasis{i}\cdot\mvec{U} \right)
\end{align}
and
\begin{align}
  \gcs\times\mvec{U}
  &=
  \frac{1}{J_c}\frac{\partial}{\partial z^i}
  \left(
  \epsilon^{ijk} U_j
  \right) \basis{k} \\
  &=
  \frac{1}{J_c}\left( \pfrac{U_3}{z^2} - \pfrac{U_2}{z^3} \right) \basis{1}
  +
  \frac{1}{J_c}\left( \pfrac{U_1}{z^3} - \pfrac{U_3}{z^1} \right) \basis{2}
  +
  \frac{1}{J_c}\left( \pfrac{U_2}{z^1} - \pfrac{U_1}{z^2} \right) \basis{3}
\end{align}
where $\mvec{U}$ is any vector field and $\epsilon^{ijk}$ is the Levi-Civita tensor.

\subsection{Gyrokinetic Equation in Field Aligned Coordinates}

The GK equation in computational coordinates becomes
\begin{align}
  \pfrac{(\mathcal{J}f)}{t} 
  + 
  \frac{1}{J_c}\frac{\partial}{\partial z^i}
  (
  J_c \mathcal{J}(\dbasis{i}\cdot\dot{\mvec{x}}) f
  )
  +
  \pfraca{v_\parallel}
  (\mathcal{J}\dot{v}_\parallel f) = 0.
  \label{eq:gk-zi}  
\end{align}
Further, in these coordinates we have
\begin{align}
  \mvec{B}^*
  =
  (\mvec{B}\cdot\dbasis{3})\basis{3}
  +
  \frac{m v_\parallel}{q} 
  \frac{1}{J_c}
  \frac{\partial}{\partial z^i}
  \left(
  \epsilon^{ijk} b_j
  \right) \basis{k}
\end{align}
where $b_j = \basis{j}\cdot\buni$.  Hence, we have
\begin{align}
  \dbasis{i}\cdot\mvec{B}^*
  =
  (\mvec{B}\cdot\dbasis{3})\delta\indices{^i_3}
  +
  \frac{m v_\parallel}{q} 
  \frac{1}{J_c}
  \frac{\partial}{\partial z^k}
  \left(
  \epsilon^{kji} b_j
  \right).
  \label{eq:em-dot-B*}
\end{align}
Further, we can compute
\begin{align}
  \dbasis{i}\cdot(\buni\times\gcs H)
  =
  \dbasis{i}\cdot(\dbasis{j}\times\dbasis{k}) b_j \pfrac{H}{z^k}
  =
  \frac{\epsilon^{ijk}}{J_c} b_j \pfrac{H}{z^k}.
\end{align}
Hence, we have
\begin{align}
  \dbasis{i}\cdot\dot{\mvec{x}}
  =
  \frac{v_\parallel}{B^*_\parallel} (\dbasis{i}\cdot\mvec{B}^*)
  +
  \frac{1}{q B^*_\parallel}
  \frac{\epsilon^{ijk}}{J_c} b_j \pfrac{H}{z^k}.
  \label{alpha config}
\end{align}
Further, we have
\begin{align}
  \dot{v}_\parallel
  =
  -\frac{(\dbasis{k}\cdot\mvec{B}^*)}{m B_\parallel^*}\pfrac{H}{z^k}.
  \label{alpha phase}
\end{align}
We can again use \eqr{\ref{eq:em-dot-B*}} to compute $\dbasis{k}\cdot\mvec{B}^*$.

The gyrokinetic Poisson equation in computational coordinates becomes
\begin{align}
  -\frac{1}{J_c}
  \frac{\partial}{\partial z^i}
  \left(
  J_c \varepsilon_\perp \dbasis{i}\cdot\nabla_\perp \phi
  \right)
  =
  \sum_s q_s \int \mathcal{J} f_s \thinspace d^3\mvec{v}.
  \label{poisson}
\end{align}
We can compute
\begin{align}
  \dbasis{i}\cdot\nabla_\perp \phi
  &=
  \dbasis{i}\cdot\dbasis{j}\pfrac{\phi}{z^j}
  -
  (\dbasis{i}\cdot \buni) (\buni\cdot\dbasis{m})\pfrac{\phi}{z^m} \\
  &=
  g^{ij} \pfrac{\phi}{z^j}
  -
  \delta\indices{^i_3}\frac{1}{\lVert \basis{3} \rVert^2 }
    \pfrac{\phi}{z^3}.
  \label{ei dot gradperp}
\end{align}
Note that in gyrokinetics we typically drop the derivatives in $z^3$ in the gyrokinetic Poisson
equation due to the assumption that gradient scale lengths in the parallel direction are much longer than those in the perpendicular direction.

\subsection{Simplifications in Axisymmetric Limit}

For divertor design, axisymmetric simulations which are 2D rather than 3D in configuration space are often used. In these simulations, 
cross-field transport is modeled with ad-hoc diffusive terms.
If we take the second computational coordinate $z^2$ as our ignorable coordinate assuming $\partial F/\partial z^2 = 0$ for all quantities $F$, we get the following equations of motion by taking $i=1,3$ in Eq.~\ref{alpha config}
\begin{equation}
\mathbf e^1 \cdot \mathbf{\dot x} = \dot z^1 = -\frac{mv_\parallel^2}{qJ_cB^*_\parallel}\pdv{b_2}{z^3} + \frac{b_2}{qJ_cB_\parallel^*}  \pdv{H}{z^3}
\label{eq:z1}
\end{equation}

\begin{equation}
\mathbf e^3 \cdot \mathbf{\dot x} =  \dot z^3 = \frac{Cv_\parallel}{J_cB_\parallel^*} + \frac{mv_\parallel^2}{qJ_cB^*_\parallel}\pdv{b_2}{z^1} + \frac{b_2}{qJ_cB_\parallel^*}  \pdv{H}{z^1}
\label{eq:z3}
\end{equation}

and Eq.~\ref{alpha phase}
\begin{equation}
\dot v_\parallel = -\frac{C}{mJ_cB_\parallel^*}\pdv{H}{z^3} + \frac{v_\parallel}{qJ_cB_\parallel^*} \left(\pdv{b_2}{z^3}\pdv{H}{z^1} - \pdv{b_2}{z^1}\pdv{H}{z^3}\right)
\label{eq:vpar}
\end{equation}

Neglecting derivatives in $z^2$ in Eq.~\ref{poisson} by gives the axisymmetric limit of the gyrokinetic poisson equation



After dropping the $z^3$ derivatives in Eq.~\ref{poisson}, as is typically justified because of the long parallel and short perpendicular wavelengths present in tokamaks, neglecting derivatives in $z^2$ gives the axisymmetric limit of the gyrokinetic poisson equation

\begin{equation}
\rho = -\frac{1}{J_c} \pdv{}{z^1} \Big[ J_c\epsilon_\perp g^{11} \pdv{\phi}{z^1} \Big]
\end{equation}

\section{Coordinate System}
\label{sec:coordinates}
\subsection{Coordinate Definitions}
Tokamak equilibrium magnetic fields are axisymmetric and can be written as~\citep{Cerfon2010}
\begin{equation}
\mathbf{B}=\frac{F(\psi)}{R} \mathbf{\hat e}_\phi+\frac{1}{R} \nabla \psi \times \mathbf{\hat e}_\phi
\end{equation}
where $\mathbf{\hat e}_\phi$ is a unit vector and $\mu_0$ is the vacuum permeability.
The poloidal flux $\psi$ will satisfy the Grad-Shafranov equation shown here in cylindrical $(R,Z,\phi)$ coordinates.
\begin{equation}
R \frac{\partial}{\partial R}\left(\frac{1}{R} \frac{\partial \psi}{\partial R}\right)+\frac{\partial^2 \psi}{\partial Z^2}=-\mu_0 R^2 \frac{d p}{d \psi}-F \frac{d F}{d \psi}
\label{eq:GS}
\end{equation}
where $F$ is the poloidal current and $p$ is the pressure. There are equilibrium codes such as the Python package FreeGS~\citep{freegs-docs, amorisco2024}, that solve Eq.~\ref{eq:GS} for $\psi(R,Z)$ and provide the solution in the commonly used G-EQDSK format~\cite{lao1997geqdsk}.

Given $\psi(R,Z)$ we choose to use field-aligned coordinates $(z^1,z^2,z^3) = (\psi,\alpha, \theta)$ where $\alpha$ is the field line label and $\theta$ is the poloidal projection of the length along the field line normalized to $2\pi$. We choose these coordinates such that our field can be represented in the Clebsch form with $C = 1$ as
\begin{equation}
\boldsymbol{B} = \nabla \psi \times \nabla \alpha
\label{clebsch}
\end{equation}

Note that there are many possible choices of the parallel coordinate $\theta$ depending on the topology one wishes to represent. For example in the core of a tokamak where flux surfaces are closed, one could choose the actual poloidal angle as the parallel coordinate $\theta$. However, as discussed in~\cite{leddy2017}, in the scrape-off layer, the actual poloidal angle is not a suitable choice because typically more than one point on the same field would have the same value of $\theta$. (In the poloidal plane, a line of constant poloidal angle will intersect the same flux surface twice in the SOL). For the outer SOL of a double-null tokamak configuration, the cylindrical coordinate $Z$ could be a suitable choice but that would of course not work for the core and would restrict the divertor plates to be horizontal in the R-Z plane which is undesirable. The choice of poloidal arc length as the parallel coordinate $\theta$ which we make here is suitable for both the open and closed field line regions of a tokamak and allows for flexible divertor plate shapes. More details on the derivation of our coordinate system can be found in~\cite{Mandell_Thesis} and other possible choices of the parallel coordinate can be found in~\cite{Jardin}.

In order to have a generalized poloidal angle that sweeps out equal poloidal arc-lengths, we choose the Jacobian to be
\begin{equation}
J_c =s(\psi) \frac{R}{|\nabla \psi|}.
\label{eq:jacexpr}
\end{equation}
Note here that the Jacobian is proportional to $1/|\nabla \psi|$. This will be true, regardless of the choice of parallel coordinate, for field aligned coordinate systems. The coordinate singularity discussed earlier results from the fact that $\nabla\psi$ vanishes at X-points and O-points, which causes the Jacobian to diverge. 

With this Jacobian, the $\theta$ coordinate, parameterized in terms of the cylindrical $Z$ coordinate, is given by
\begin{equation}
\theta(R, Z)=\frac{1}{s(\psi(R, Z))} \int_{Z_{\text {lower }}(\psi)}^{Z(\psi)} \sqrt{1+\left(\frac{\partial R\left(\psi, Z^{\prime}\right)}{\partial Z^{\prime}}\right)^2} \mathrm{~d} Z^{\prime} - \pi
\label{eq:theta}
\end{equation}
where the normalization factor is 
\begin{equation}
s(\psi)=\frac{1}{2 \pi} \oint \mathrm{d} \ell_p=\frac{1}{2\pi} \int_{Z_{\text {lower }}(\psi)}^{Z_{\text {upper }}(\psi)} \sqrt{1+\left(\frac{\partial R\left(\psi, Z^{\prime}\right)}{\partial Z^{\prime}}\right)^2} \mathrm{~d} Z^{\prime}
\label{eq:normalization}
\end{equation}
Now we define the last coordinate such that Eq.~\ref{clebsch} is satisfied
\begin{equation}
\alpha(R, Z, \phi)=-\phi+F(\psi) \int_{Z_{\text{lower }}(\psi)}^{Z(\psi)} \frac{1}{|\nabla \psi| R\left(\psi, Z^{\prime}\right)} \sqrt{1+\left(\frac{\partial R\left(\psi, Z^{\prime}\right)}{\partial Z^{\prime}}\right)^2} \mathrm{~d} Z^{\prime}
\label{eq:alpha}
\end{equation}
where $F(\psi) = RB_\phi$.

We make the following choice of computational coordinates: $(z^1, z^2, z^3) = (\psi, \alpha, \theta)$. This choice along with Eqs.~\ref{eq:normalization},~\ref{eq:theta}, and ~\ref{eq:alpha} define the mapping of computational coordinates $(\psi, \alpha, \theta)$ to physical $(R,Z,\phi)$ coordinates where $\theta \in [-\pi, \pi]$ and $\alpha \in [-\pi, \pi]$. From the mapping we can compute tangent vectors, dual vectors, and then metric coefficients, which are written explicitly in Eqs.~\ref{eq:metrics1}-~\ref{eq:metrics6}.

\begin{figure}
    \subfloat[\label{fig:dnarrows}
    Schematic for field line tracing for a double null tokamak configuration. There are 5 distinct regions: the outboard scrape-off-layer, the inboard scrape-off-layer, the lower private flux region, the upper private region, and the core. We plot one flux surface in black in each region. The tracing of the flux surface starts in each region at $Z_{lower}(\psi)$  (marked in blue) and stops at $Z_{upper}(\psi)$ (marked in red) in accordance with Eq.~\ref{eq:normalization}. The green arrows indicate the direction of the tracing in each region.]{
    \includegraphics[width=0.5\linewidth, valign=t]{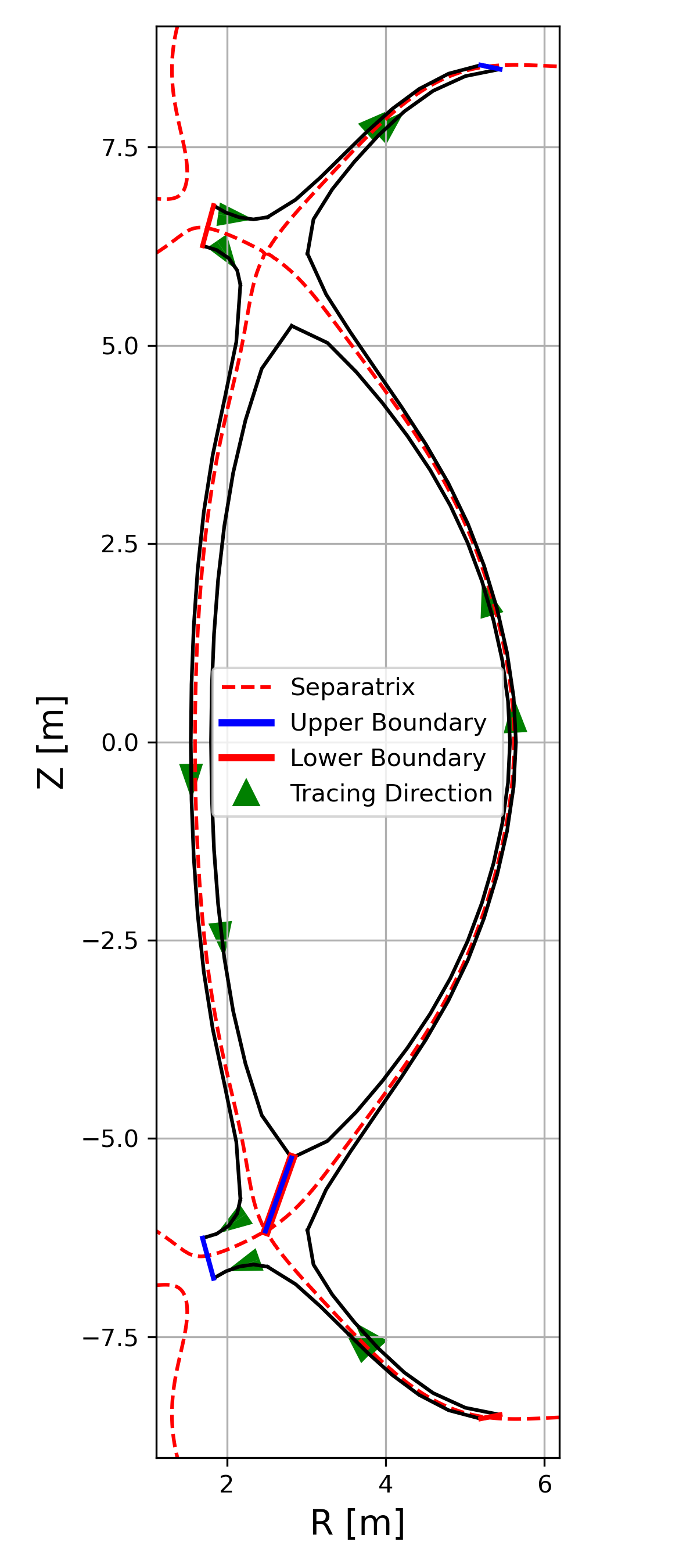}
    }
    \subfloat[\label{fig:snarrows}
    Schematic for field line tracing for a single null tokamak configuration. There are 3 distinct regions: the scrape-off-layer, the private flux region, and the core. We plot one flux surface in black in each region. The tracing of the flux surface starts in each region at $Z_{lower}(\psi)$  (marked in red) and stops at $Z_{upper}(\psi)$ (marked in blue) in accordance with Eq.~\ref{eq:normalization}. The green arrows indicate the direction of the tracing in each region.]{
    \includegraphics[width=0.5\linewidth, valign=t]{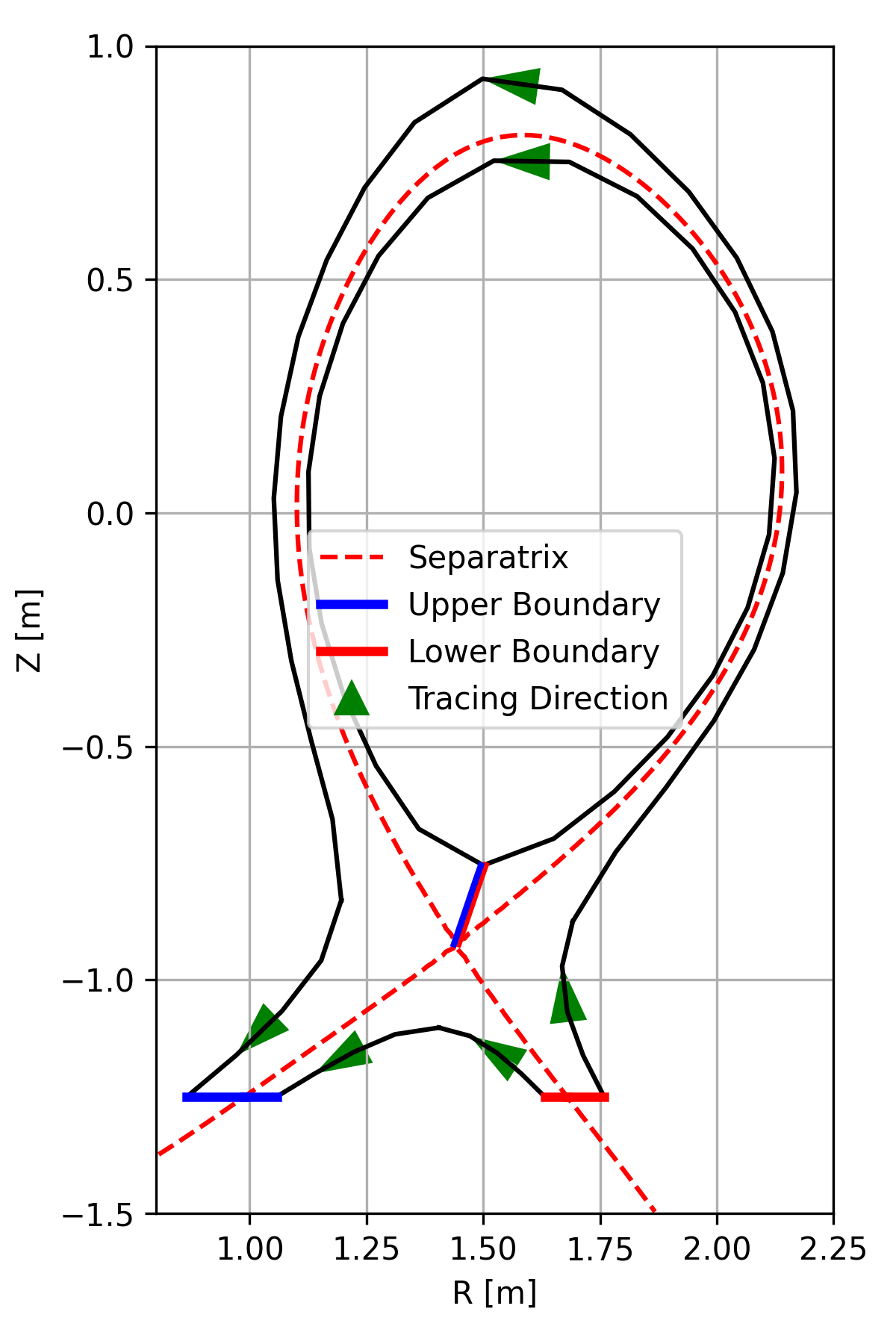}
    }
    \caption{Schematic for field line tracing in a double null (a) and single null (b) configuration.
    \label{fig:arrows}}
\end{figure}

The integrals in ~\ref{eq:theta},~\ref{eq:normalization}, and~\ref{eq:alpha} are along contours of constant $\psi$. The Z limits of the integration can be chosen based on the part of the poloidal plane one wishes to trace. For example integral in Eq.~\ref{eq:normalization} traces from divertor plate to divertor plate in the SOL but makes a complete poloidal circuit in the core. For a double null tokamak configuration there are 5 distinct topological regions: the outboard SOL, the inboard SOL, the lower private flux region, the upper private region, and the core. In Fig.~\ref{fig:dnarrows} we show how each region is traced for a double null tokamak.
For a single null tokamak configuration there are 3 distinct topological regions: the SOL, the private flux region, and the core. In Fig.~\ref{fig:snarrows} we show how each region is traced for a lower single null tokamak.

\section{Discretization of the Gyrokinetic Equation: Avoiding the X-point }
\label{sec:discretization}

The gyrokinetic equation, Eq.~\ref{eq:gk-zi} for the evolution of $F = \mathcal{J} J_c f$ in the axisymmetric limit becomes
\begin{equation}
\pdv{F}{t} + \dot{z}^1\pdv{F}{z^1} + \dot{z}^3\pdv{F}{z^3} + \dot{v_\parallel}\pdv{F}{v_\parallel} = 0.
\label{eq:gk_axi}
\end{equation}

We use a Discontinuous Galerkin (DG) scheme to discretize this equation as described in ~\cite{Mana25, Ammar2019, Mandell2020}. The discrete approximation of $F$ in each cell $K_i$ is given by
\begin{equation}
F_i=\sum_{k=1}^{N_b} F_i^{(k)} \psi_i^{(k)}
\label{eq:expansion}
\end{equation}
where $\psi_i$ are the phase-space basis functions and $N_b$ is the number of basis functions.
The discrete form of Eq.~\ref{eq:gk_axi} can be obtained by projecting it onto the phase space basis $\psi_j^{(k)}$ in cell $K_j$ and integrating by parts
\begin{equation}
\begin{aligned}
& \int_{K_j} d\mathbf{z} dv_\parallel d\mu \psi_j^{(\ell)} \frac{\partial F}{\partial t}+\oint_{\partial K_j} \mathrm{~d} \mathbf{S}_{\mathrm{i}} dv_\parallel d\mu \psi_{j \pm}^{(\ell)} \dot{z}_{ \pm}^i \widehat{F}_{ \pm}+\oint_{\partial K_j} d \mathbf{z} d\mu \psi_{j \pm}^{(\ell)} \dot{v}_{\| \pm} \widehat{F_{ \pm}} \\
& \quad-\int_{K_j} d\mathbf{z} dv_\parallel d\mu\left(\frac{\partial \psi_j^{(\ell)}}{\partial z^i} \dot{z}^i+\frac{\partial \psi_j^{(\ell)}}{\partial v_\parallel} \dot{v}_{\|}\right) F=0 .
\end{aligned}
\end{equation}
where $d\mathbf S_i$ is the surface element perpendicular to the i-th direction and $\widehat{F_\pm}$ is the upwind flux evaluated at the upper and lower edge of the cell in direction i (see appendix~\ref{sec:appendix} for more details about numerical fluxes).

Substituting in the expansion of $F$ in the first term of Eq.~\ref{eq:expansion} and making use of the orthonormality relation $\int_{K_j} d\mathbf z dv_\parallel d\mu \psi_j^{(\ell)}\psi_j^{(k)} = \delta_{lk}$, we get the time evolution of each expansion coefficient of $F$
\begin{equation}
\begin{aligned}
 &\frac{\partial F_j^{(\ell)}}{\partial t}+\oint_{\partial K_j} \mathrm{~d} \mathbf{S}_{\mathrm{i}} dv_\parallel d\mu \psi_{j \pm}^{(\ell)} \dot{z}_{ \pm}^i \widehat{F}_{ \pm}+\oint_{\partial K_j} d \mathbf{z} d\mu \psi_{j \pm}^{(\ell)} \dot{v}_{\| \pm} \widehat{F_{ \pm}} \\
 &\quad-\int_{K_j} d\mathbf{z} dv_\parallel d\mu\left(\frac{\partial \psi_j^{(\ell)}}{\partial z^i} \dot{z}^i+\frac{\partial \psi_j^{(\ell)}}{\partial v_\parallel} \dot{v}_{\|}\right) F=0 .
 \label{eq:evolution}
\end{aligned}
\end{equation}

We evaluate the integrals in Eq.~\ref{eq:evolution} analytically using DG expansions of the characteristics  $\dot z^i$ and $\dot v_\parallel$ on the phase basis in the volume term (the fourth term) and DG expansions of the fluxes ($\dot{z}^i\hat{F}_\pm$ and $\dot{v_\parallel}\hat{F}_\pm$) in the second and third terms (the surface terms). In the last term (the volume term) the expansion of the characteristic velocities ($\dot{z}^i$ and $\dot{v_\parallel}$) are constructed by evaluating the characteristics at interior Gauss-Legendre quadrature points and converting to a modal representation. In the second and third terms, the expansion of the fluxes are calculated by evaluating the flux at surface Gauss-Legendre quadrature points and converting to a modal representation. 
Labeling the characteristics and fluxes based on whether they are calculated by evaluation at interior or surface quadrature points with a subscripts $int$ and $surf$ respectively, we can rewrite Eq.~\ref{eq:evolution} as

\begin{equation}
\begin{aligned}
 &\frac{\partial F_j^{(\ell)}}{\partial t}+\oint_{\partial K_j} \mathrm{~d} \mathbf{S}_{\mathrm{i}} dv_\parallel d\mu \psi_{j \pm}^{(\ell)} (\dot{z}^i \widehat{F})_{\pm , surf} +\oint_{\partial K_j} d \mathbf{z} d\mu \psi_{j \pm}^{(\ell)} (\dot{v}_{\|} \widehat{F})_{\pm , surf} \\
 &\quad-\int_{K_j} d\mathbf{z} dv_\parallel d\mu\left(\frac{\partial \psi_j^{(\ell)}}{\partial z^i} \dot{z}^i_{int}+\frac{\partial \psi_j^{(\ell)}}{\partial v_\parallel} \dot{v}_{\|,int}\right) F=0 .
 \label{eq:evolution_quad}
\end{aligned}
\end{equation}

\begin{figure}
    \centering
    \subfloat[
    Gauss-Legendre quadrature points on the surface (blue and green) and interior (red) points of a computational cell along with cell corners (orange). 
    \label{fig:quad_points}
    ]{
    \includegraphics[width=0.5\linewidth]{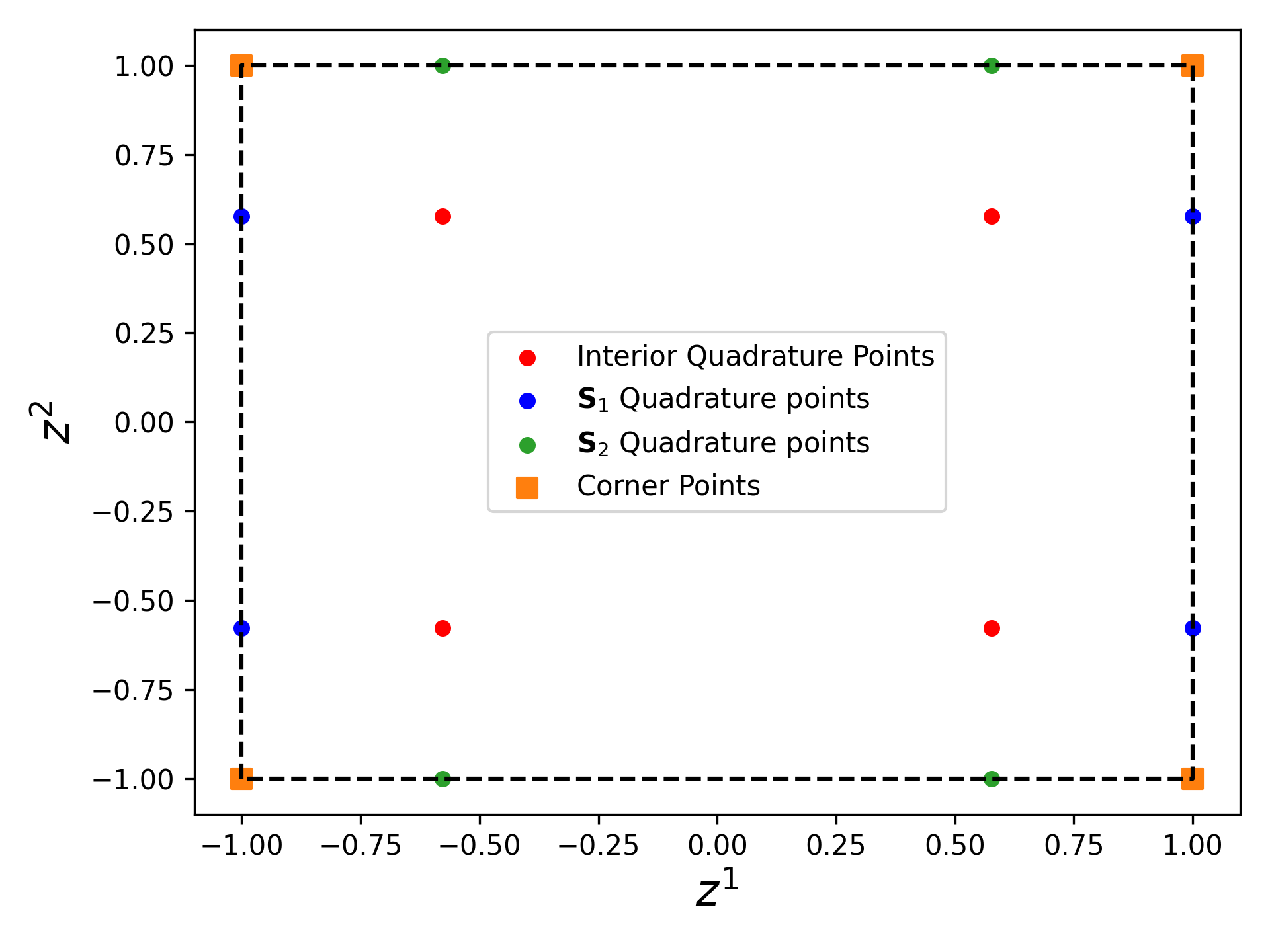}
    }
    \subfloat[
    Gauss-Legendre quadrature points on the surface (blue and green) and interior (red) points along with cell corners (orange) mapped to physical cells abutting the X-point. 
    \label{fig:xpt_quad_points}
    ]{
    \includegraphics[width=0.5\linewidth]{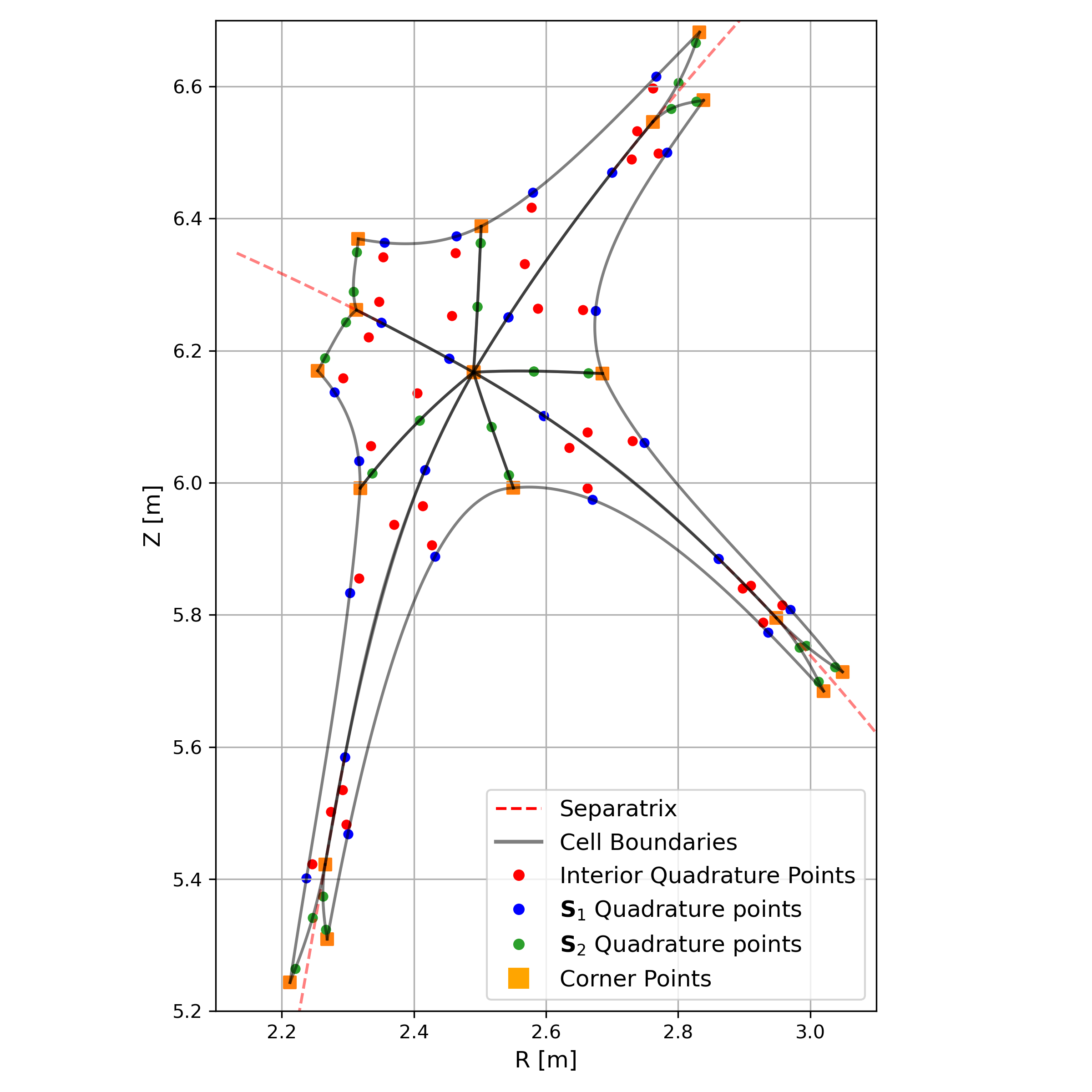}
    }
    \caption{In (a) we show the interior, surface, and corner points on the unit cell. In (b) we show these points mapped to the physical domain for cells abutting the X-point. The cell in physical space is \emph{not rectuangluar}, allowing for an accurate representation of the flux-surface geometry.  The surface and interior nodes used for the evaluation of geometric quantities do not lie directly on the X-point and are thus well defined. 
    \label{fig:xpt_quad_points_mapped}
    }
\end{figure}

An example of the quadrature points used in 2D is depicted in Fig.~\ref{fig:quad_points}. For example to construct the volume representation $\dot{z^i_{int}}$ in this cell, we evaluate $\dot z^i$ at the 4 red points and convert to a modal representation. To calculate the surface representation $(\dot z^1\hat{F})_{+,surf}$ at the upper $z^1$ edge of this cell we would evaluate $\dot z^1\hat{F}$ at the two blue points at $z^1=1$ and convert to a modal representation.
The use of an orthonormal, modal representation for the DG fields allows us to significantly reduce the computational cost of DG~\citep{HakimandJuno2020} while respecting the need to eliminate aliasing errors in DG discretizations of kinetic equations~\citep{Juno18}.

This method of evaluating the characteristics is the key feature of our algorithm that allows us to simulate magnetic geometries with an X-point as shown in Fig.~\ref{fig:xpt_quad_points_mapped}. The geometric quantities such as the Jacobian, $J_c$, contained in the characteristics written in Eqs.~\ref{eq:z1},~\ref{eq:z3}, and~\ref{eq:vpar} diverge at the X-point as mentioned below Eq.~\ref{eq:jacexpr}. However, since we evaluate the characteristics at either interior quadrature points or surface quadrature points and not corner points, we can avoid evaluating any geometric quantities at the X-point as long as cell corners lie at the X-point, which our multi-block grid generation routine ensures. 
The gyrokinetic poisson equation, Eq.~\ref{poisson}, also benefits from the distinction between corner and interior evaluations. Our solution, described in detail in ~\cite{Mana25}, makes use of the interior geometric quantities to avoid the coordinate singularity. The requirements for consistent surface fluxes at boundaries between blocks with different mappings is described in appendix~\ref{sec:appendix}.

\section{Grid Generation and Geometric Quantities}
\label{sec:computation}

In order to conduct simulations, we need to generate a physical simulation grid and then calculate all of the geometric quantities appearing in the equations of motion on that grid. All of the geometric quantities required can be extracted from two basic quantities: the magnitude of the magnetic field $B(\psi, \alpha, \theta)$ and the tangent vectors defined by
\begin{equation}
\mathbf{e}_i = \pdv{\mathbf x}{z^j}
\label{tangent}
\end{equation}
where $\mathbf x = (x,y,z)$ are the global cartesian coordinates and $(z^1, z^2, z^3) = (\psi, \alpha, \theta)$ are the computational coordinates.

\subsection{Representation of Magnetic Field}
The starting point for our grid generation is a tokamak equilibirum provided by the commonly used G-EQDSK format~\citep{lao1997geqdsk}. The G-EQDSK format provides $\psi(R,Z)$ on an $N_R \times N_Z$ grid and from that we can construct a DG expansion of $\psi(R,Z)$ on the same grid. 
G-EQDSK files also give the toroidal magnetic field by providing $F(\psi) = RB_\phi$ on a grid of length $N$ from which we can construct a DG expansion of $F(\psi)$. 
We use either a biquadratric or bicubic representation of $\psi(R,Z)$ for the field line tracing described in section~\ref{sec:exact_mapping} and for calculating the magnitude of the magnetic field at each grid point. The biquadratic representation offers a speedup over the bicubic representation in the grid generation process because it enables a simple and fast root finding procedure. The magnetic field components and magnitude can be calculated from $\psi(R,Z)$ and $F(\psi)$ in cylindrical coordinates as
\begin{subequations}
\begin{align}
B_R & =\frac{1}{R} \frac{\partial \psi}{\partial Z}=\frac{\partial}{\partial Z}\left(\frac{\psi}{R}\right) \label{eq:BR}\\
B_Z & = -\frac{1}{R} \frac{\partial \psi}{\partial R}=-\frac{\partial}{\partial R}\left(\frac{\psi}{R}\right)-\frac{\psi}{R^2} \label{eq:BZ}\\
B_\phi &= \frac{F(\psi)}{R}\\
B &= \|\mathbf B \| = \sqrt{B_R^2+B_Z^2+B_\phi^2}.
\end{align}
\end{subequations}

The poloidal magnetic field is $\mathbf B_{pol} = B_R \hat{\mathbf R} + B_Z\hat{\mathbf Z}$. Looking at Eq.~\ref{eq:jacexpr} and Eqs.~\ref{eq:BR} and~\ref{eq:BZ}, one can now see the connection between a vanishing poloidal field and the coordinate singularity at the X-point; when $\mathbf B_{pol}$ vanishes, $|\nabla\psi| = 0$, and the Jacobian, $J_c$, diverges.

\subsection{Grid Generation Algorithm}
\label{sec:exact_mapping}

We use a rectangular computational grid with extents $(L_\psi, L_\alpha, L_\theta)$ and number of cells $(N_\psi, N_\alpha, N_\theta)$. The grid spacing is $(\Delta\psi, \Delta\alpha, \Delta\theta) = (L_\psi/N_\psi, L_\alpha/N_\alpha, L_\theta/N_\theta)$. This computational grid has $(N_\psi+1)(N_\alpha+1)(N_\theta+1)$ nodes. 

In order to lay out a physical grid for our simulation and to calculate the geometric factors appearing in Eq.~\ref{alpha config} and Eq.~\ref{alpha phase} we calculate mapping $\mathbf x(\psi, \alpha, \theta)$ at each point on our computational grid. For each point, $(\psi_0, \alpha_0, \theta_0)$, on our grid, we calculate the mapping using the following algorithm:

\begin{itemize}
    \item {\bf Step 1:} Pick an initial $Z$ and find $R$ such that $\psi(R,Z) = \psi_0$. In practice this is done by inverting our piecewise polynomial representation of $\psi(R,Z)$ to get a polynomial $R(\psi,Z)$.
    \item {\bf Step 2:} Calculate $\theta(R,Z)$ using Eq.~\ref{eq:theta}. The integral is done with a double exponential method~\citep{Bailey} and will require doing Step 1 and evaluating the derivative of the polynomial $R(\psi,Z)$ at each quadrature point to remain on the flux surface.
    \item {\bf Step 3:} Repeat Steps 1 and 2 choosing $Z$ using a root-finder (we use Ridders method~\citep{Ridders}) until we find $R$ and $Z$ such that $\theta(R,Z) = \theta_0$.
    \item {\bf Step 4:} Calculate $\phi$ using Eq.~\ref{eq:alpha}.
    \item {\bf Step 5:} Calculate the Cartesian coordinates from the cylindrical coordinates: $X = R\cos\phi, Y=R\sin\phi$, $Z=Z$.
\end{itemize}

\subsection{Multi-Block Grids}
\label{sec:grids}
\begin{figure}
    \subfloat[\label{fig:stepgridfull}]{
    \includegraphics[width=0.55\textwidth, valign=t]{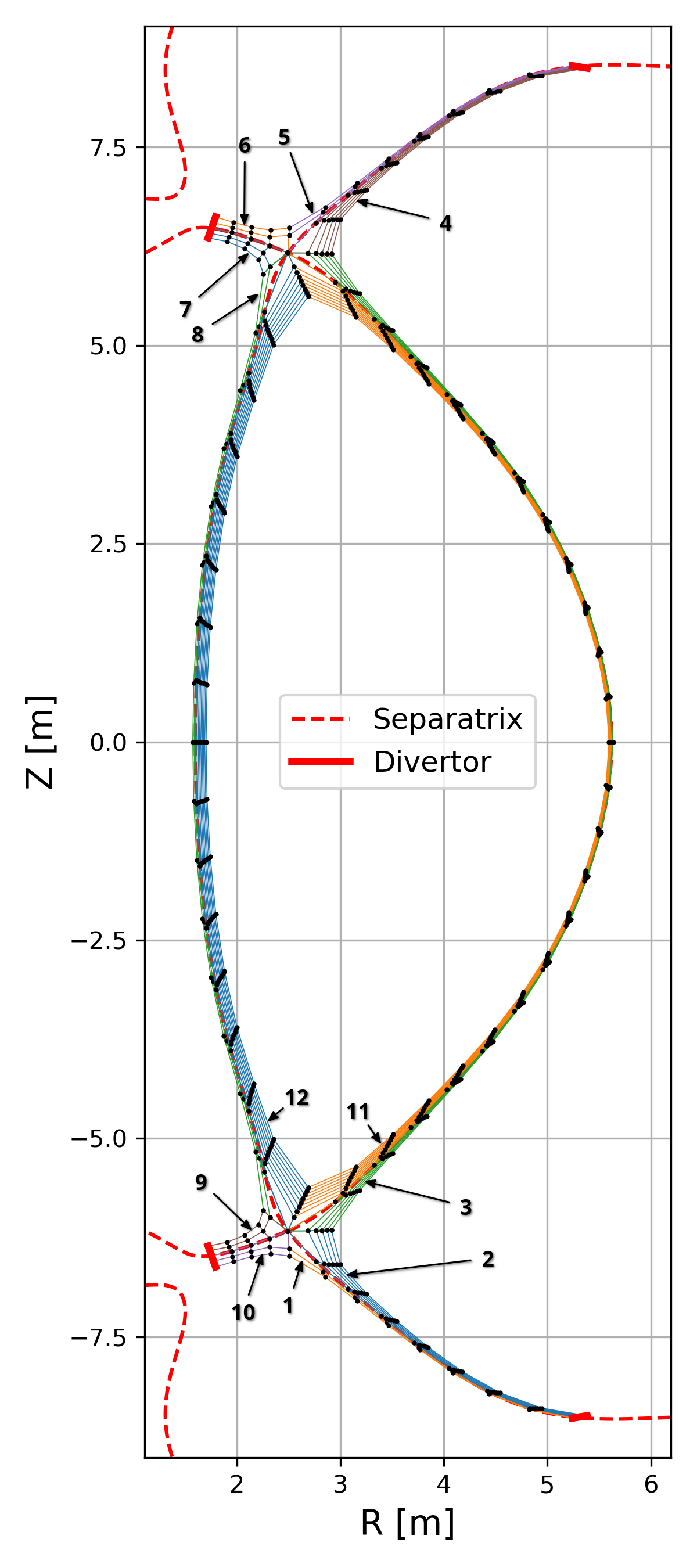}
    }
    \subfloat[\label{fig:stepgridzoom}]{
    \includegraphics[width=0.45\textwidth, valign=t]{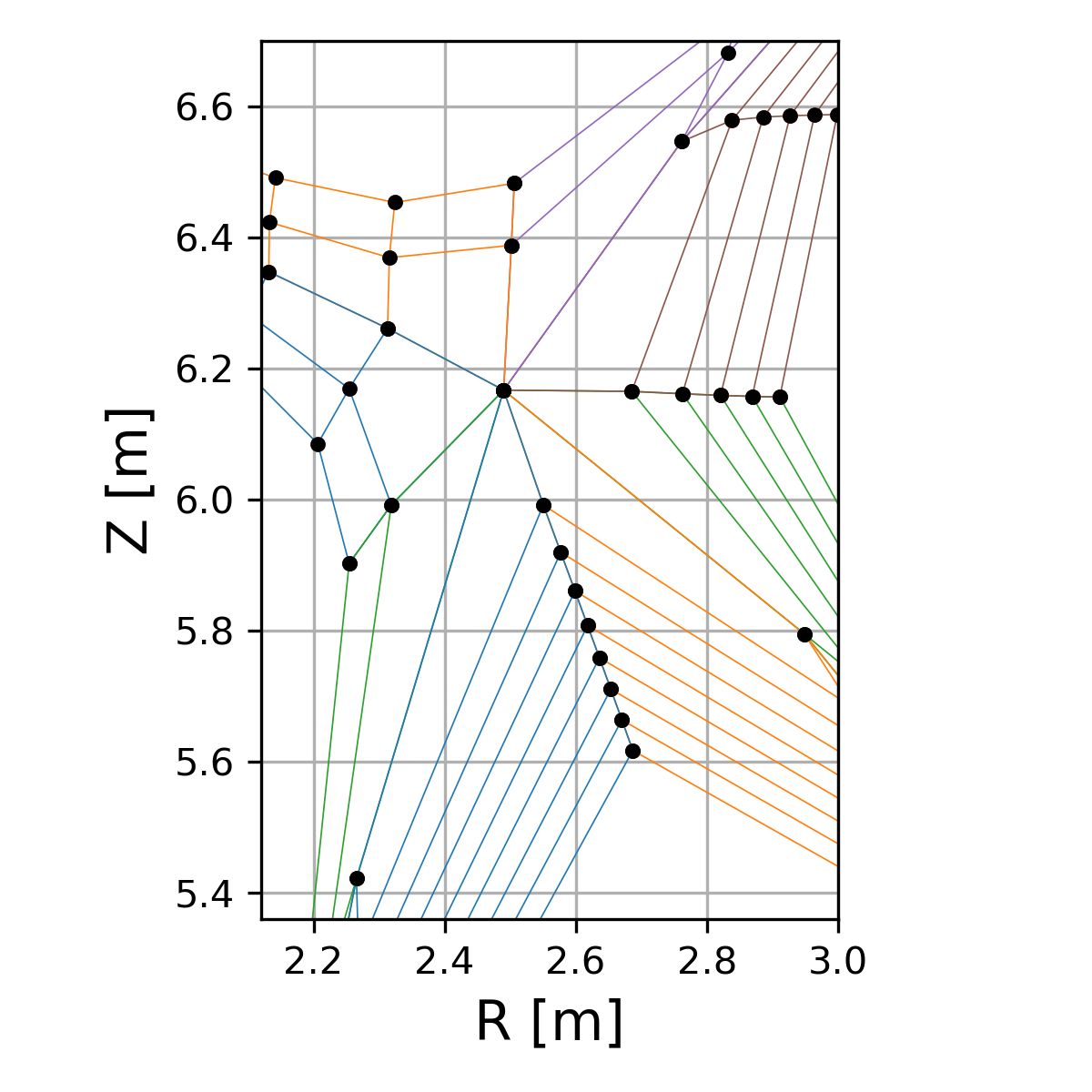}
    }\\
    \begin{flushright}
    \vspace{-10cm}
    \subfloat[\label{fig:stepgridzoomdivertor}]{ 
    \includegraphics[width=0.4\textwidth]{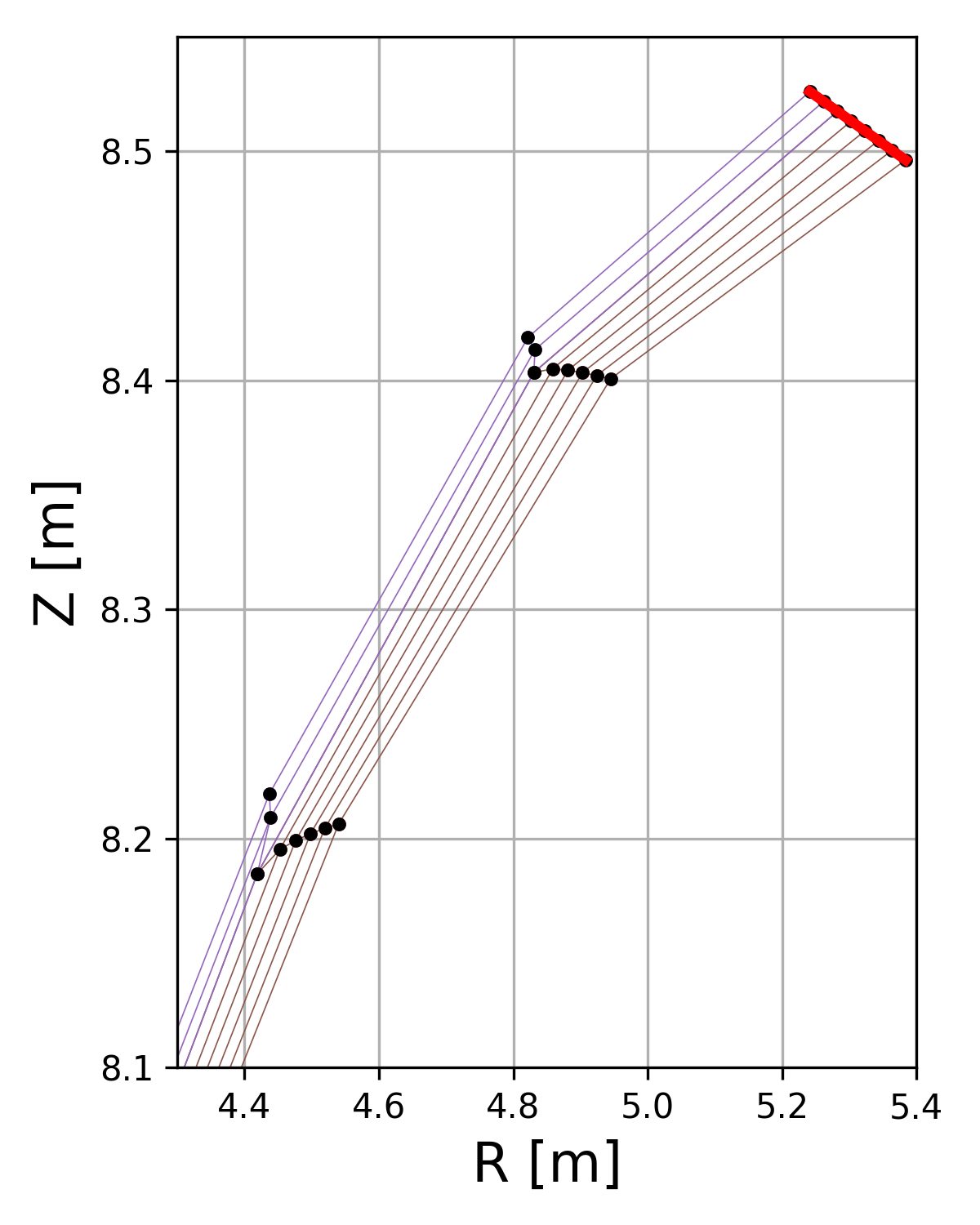}
    }
    \vspace{3cm}
    \end{flushright}
    \caption{Block layout and grid for the Spherical Tokamak for Energy Production in a double null configuration with different colors indicating different blocks and a number 1-12 labeling each block. The full grid is shown in (a), (b) shows a close-up of the grid near the upper X-point, and (c) shows a close-up of the grid near the upper outer divertor plate (red). 
    \label{fig:stepgrid} }
\end{figure}

\begin{figure}
    \subfloat[\label{fig:asdexgridfull}]{
    \includegraphics[width=0.45\textwidth, valign=t]{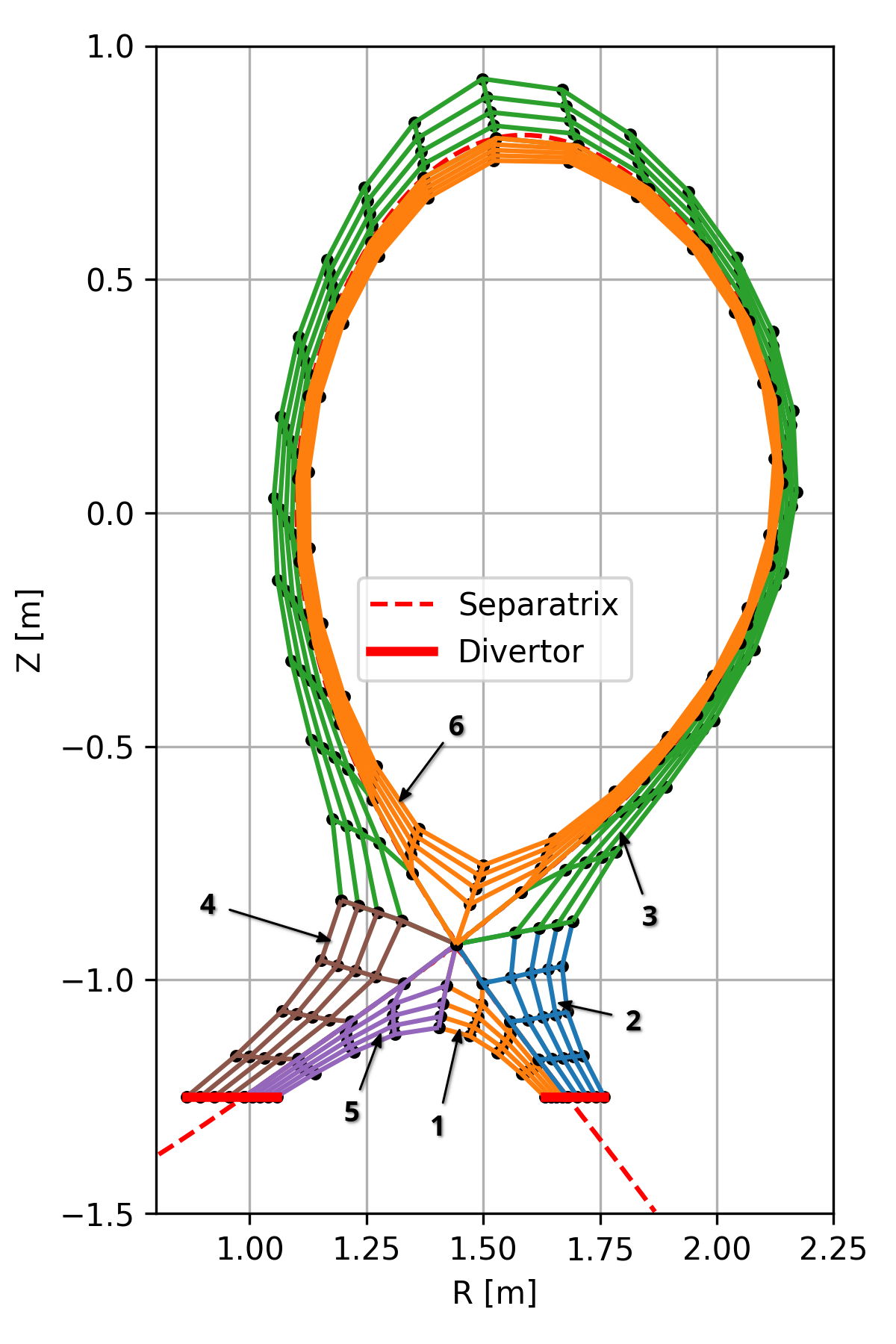}
    }
    \subfloat[\label{fig:asdexgridzoom}]{
    \includegraphics[width=0.45\textwidth, valign=t]{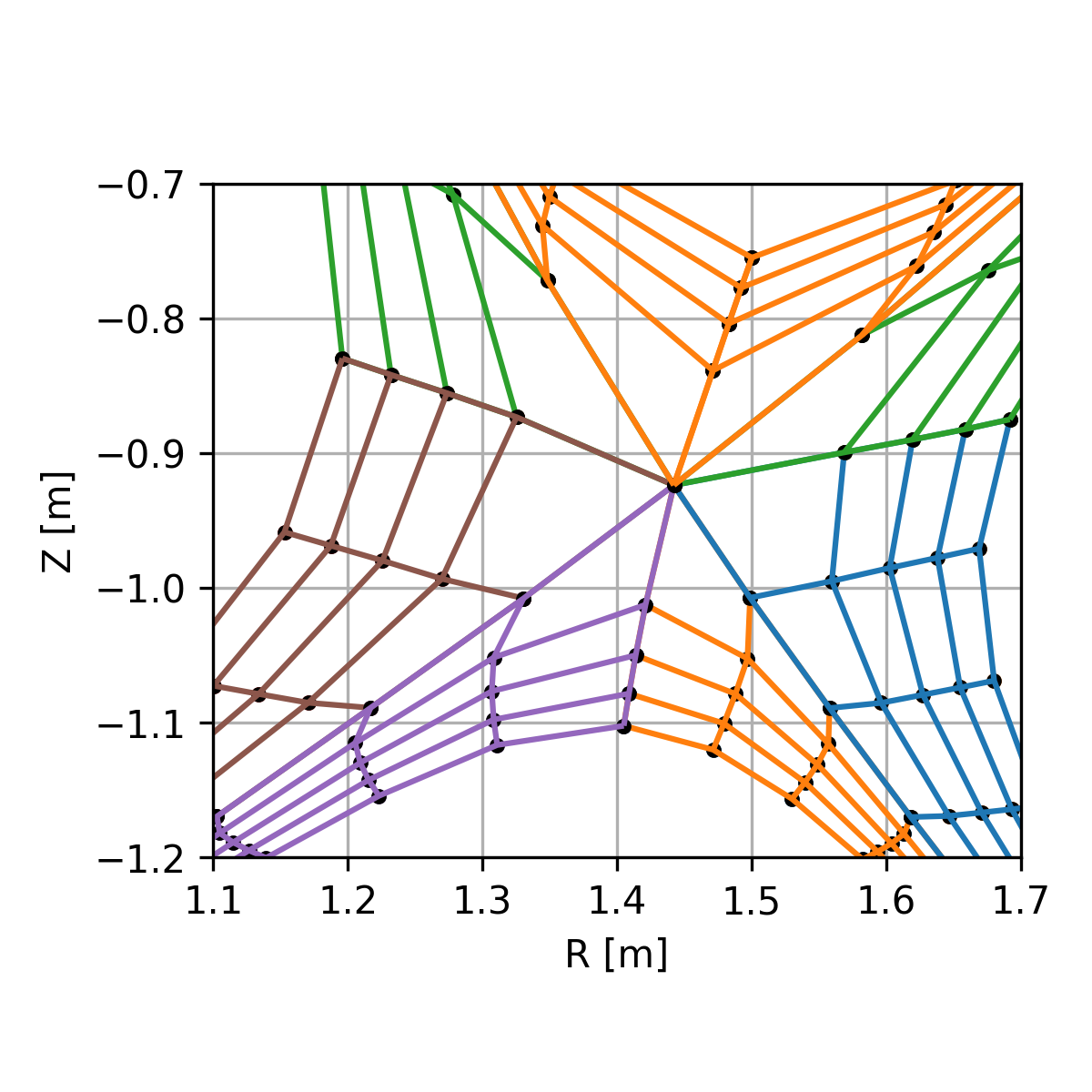}
    }
    \caption{Grid for ASDEX-Upgrade in a single null configuration with different colors indicating different blocks and a number 1-6 labeling each block. The full grid is shown in (a) and (b) shows a close-up of the grid near the X-point. 
    \label{fig:asdexgrid} }
\end{figure}

To enable simulations of domains including the core, private flux, and SOL, we break the domain up into blocks. We first break the domain up into the distinct topological regions (5 for double null and 3 for single null) described in Sec.~\ref{sec:coordinates} and then split each region at the X-point. As shown in Fig.~\ref{fig:stepgridfull}, a double null tokamak has 12 blocks each of which has one edge along the separatrix and at least one corner at the X-point. Fig.~\ref{fig:asdexgridfull} shows a lower single null tokamak with 6 blocks.

Once the domain has been split into blocks, we can generate a uniform computational grid within each block. In Fig.~\ref{fig:stepgrid} we show the multi-block grid generated for a double null configuration of STEP, and in Fig.~\ref{fig:asdexgrid} we show the multi-block grid generated for ASDEX-Upgrade~\citep{Stroth2022} in a lower single null configuration. 

\subsection{Metric Coefficients}
One we have generated mapping from computational to physical coordinates at all of the grid nodes, we can calculate the metric coefficients associated with this coordinate transformation. The tangent vectors are defined by Eq.~\ref{tangent}. The metric coefficients of the transformation are then given by
\begin{align}
    g_{ij}&= \mathbf e_i \cdot \mathbf e_j \notag \\
          &= \frac{\partial \mathbf{x}}{\partial z^i} \cdot \frac{\partial \mathbf{x}}{\partial z^j} =\sum_{k=1}^3 \frac{\partial x_k}{\partial z^i} \frac{\partial x_k}{\partial z^j}
\end{align}

Using the definitions of transformation from cartesian coordinates $(x,y,z)$ to cylindrical coordinates $(R, Z, \phi)$ 
\begin{align}
x &= R(\psi, \theta)\cos\phi(\psi, \theta, \alpha)\\
y &= R(\psi, \theta)\sin\phi(\psi, \theta, \alpha)\\
z &= Z(\psi, \theta)
\end{align}
we can express the metric coefficients, $g_{ij}$, in terms of the derivatives of the cylindrical coordinates $(R, Z, \phi)$ with respect to the computation coordinates $(\psi, \alpha, \theta)$ as follows
\begin{align}
g_{11} &= \Big(\pdv{R}{\psi}\Big)^2 + R^2\Big(\pdv{\phi}{\psi}\Big)^2 + \Big(\pdv{Z}{\psi}\Big)^2 \\
\label{eq:metrics1}
g_{12} &= R^2 \pdv{\phi}{\psi} \\
g_{13} &= \pdv{R}{\psi}\pdv{R}{\theta} + R^2\pdv{\phi}{\psi}\pdv{\phi}{\theta} + \pdv{Z}{\psi}\pdv{Z}{\theta} \\
g_{22} &= R^2 \\
g_{23}  &= R^2\pdv{\phi}{\theta} \\
g_{33} &= \Big(\pdv{R}{\theta}\Big)^2 + R^2\Big(\pdv{\phi}{\theta}\Big)^2 + \Big(\pdv{Z}{\theta}\Big)^2 .
\label{eq:metrics6}
\end{align}

We can also express the Jacobian of the coordinate transformation, $J_c$, in terms of these derivatives as follows
\begin{align}
J_c^2 &= g_{11} (g_{22} g_{33} - g_{23} g_{23}) - g_{12} (g_{12} g_{33} - g_{23} g_{13})+g_{13} (g_{12} g_{23} - g_{22} g_{13}) \notag \\
    &=  R^2 \Big[ \Big(\pdv{R}{\psi} \Big)^2\Big(\pdv{Z}{\theta} \Big)^2 +\Big(\pdv{R}{\theta} \Big)^2\Big(\pdv{Z}{\psi}\Big)^2 - 2\pdv{R}{\psi}\pdv{R}{\theta}\pdv{Z}{\psi}\pdv{Z}{\theta} \Big]\\
J_c &= R\Big[ \pdv{R}{\psi}\pdv{Z}{\theta} - \pdv{R}{\theta}\pdv{Z}{\psi}\Big].
\end{align}



The derivatives $(R,Z,\phi)$ with respect to the computational coordinate $\theta$ can be calculated directly from our representation of $\psi(R,Z)$ and the derivatives with respect to $\alpha$ are trivial:

\begin{align}
\pdv{R}{\theta} &= \sin\Big[\arctan(\pdv{R}{Z})\Big] s(\psi)\\
\pdv{Z}{\theta} &= \cos\Big[\arctan(\pdv{R}{Z})\Big] s(\psi)\\
\pdv{\phi}{\theta} &= \frac{F(\psi)}{R|\grad\psi|} s(\psi) \\
\pdv{R}{\alpha} &= 0 \\
\pdv{Z}{\alpha} &= 0 \\
\pdv{\phi}{\alpha} &= 1
\label{exact_derivs}
\end{align}
The derivative $\pdv{R}{Z}$ appearing in the derivatives with respect to $\theta$ can be calculated easily from our biquadratic representation of $\psi(R,Z)$. To calculate the remaining 3 derivatives with respect to $\psi$, $\pdv{R}{\psi}$, $\pdv{Z}{\psi}$, and $\pdv{\phi}{\psi}$, we use second order finite differences.



%

\section{Test Case: STEP Simulation}
\label{sec:example}

\begin{figure}
    \subfloat[\label{fig:step_density}]{
    \includegraphics[width=0.33\textwidth, valign=t]{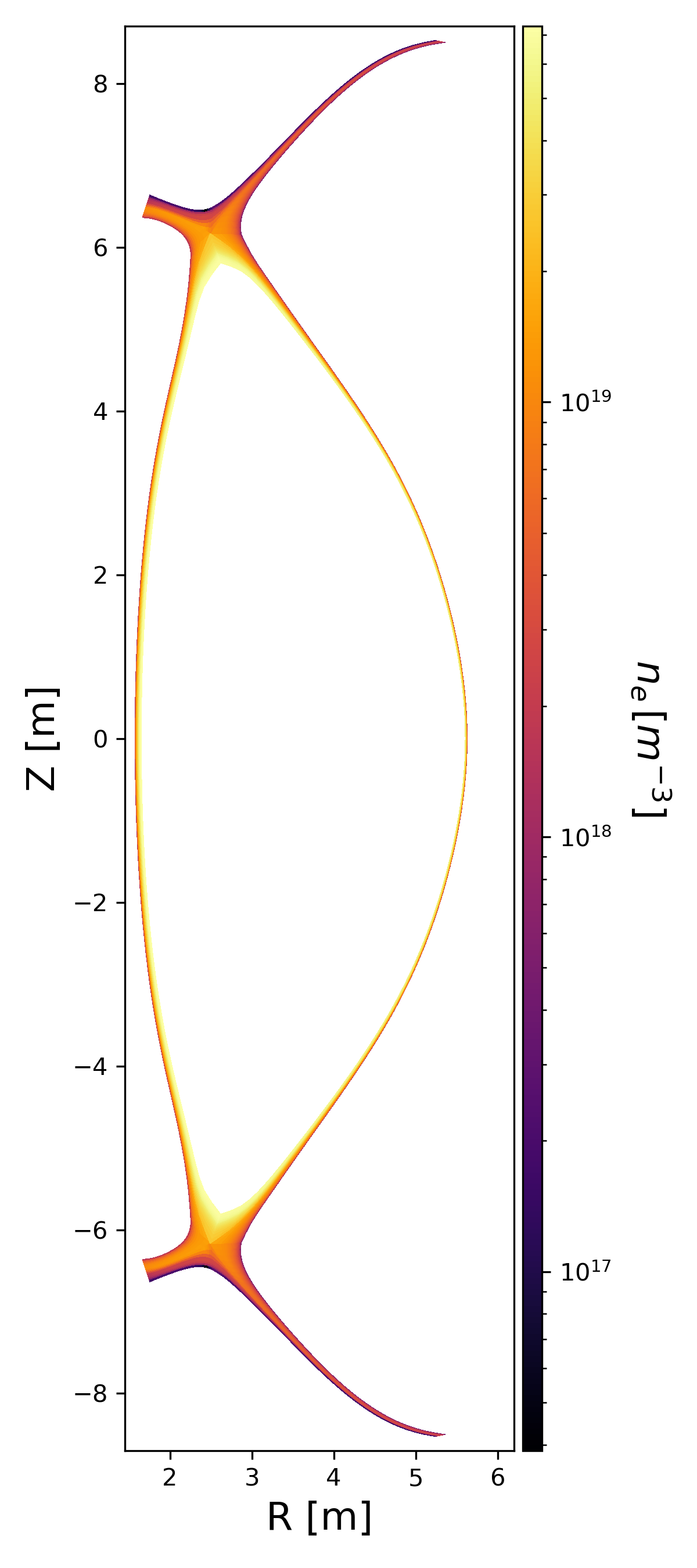}
    }
    \subfloat[\label{fig:step_temp}]{
    \includegraphics[width=0.33\textwidth, valign=t]{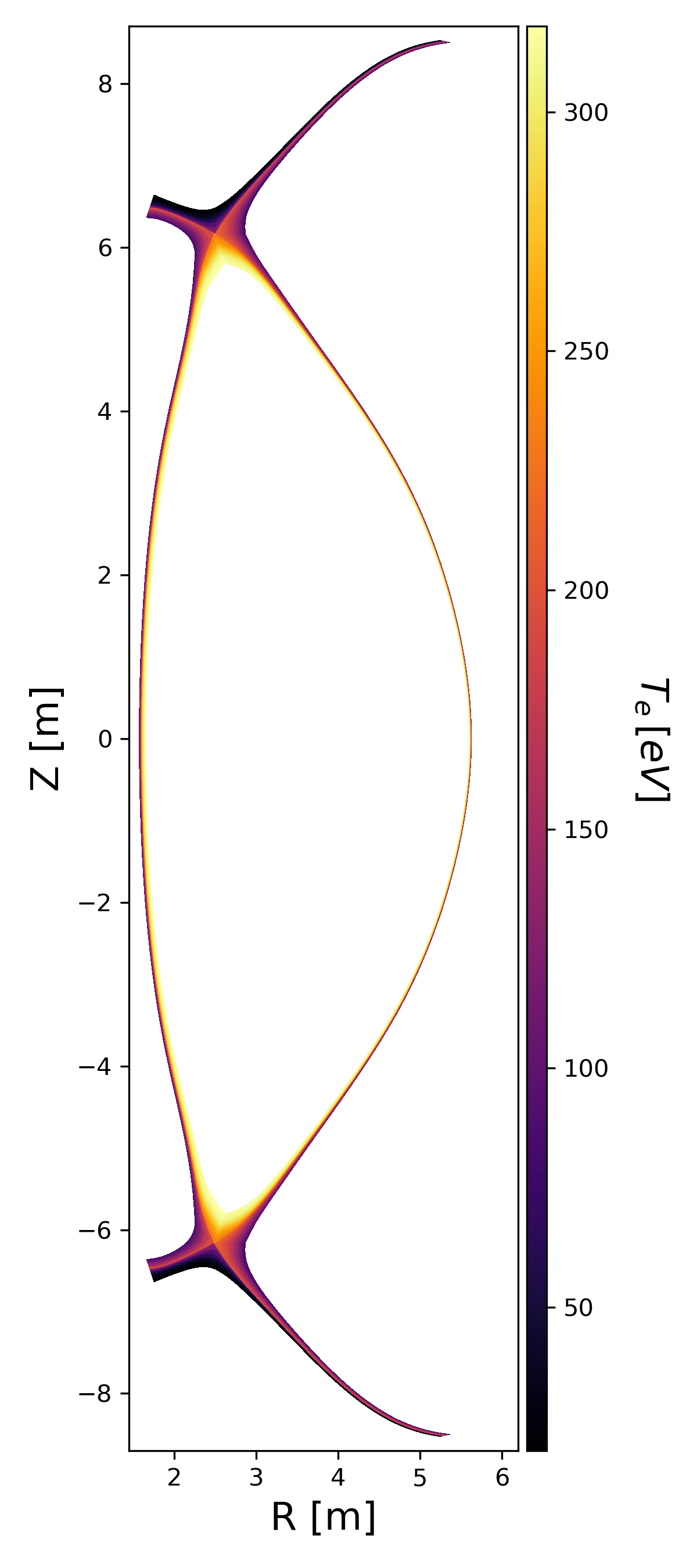}
    }
    \vspace{-11cm}
    \begin{flushright}
    \subfloat[\label{fig:step_densityzoom}]{
        \includegraphics[width=0.33\textwidth, valign=t]{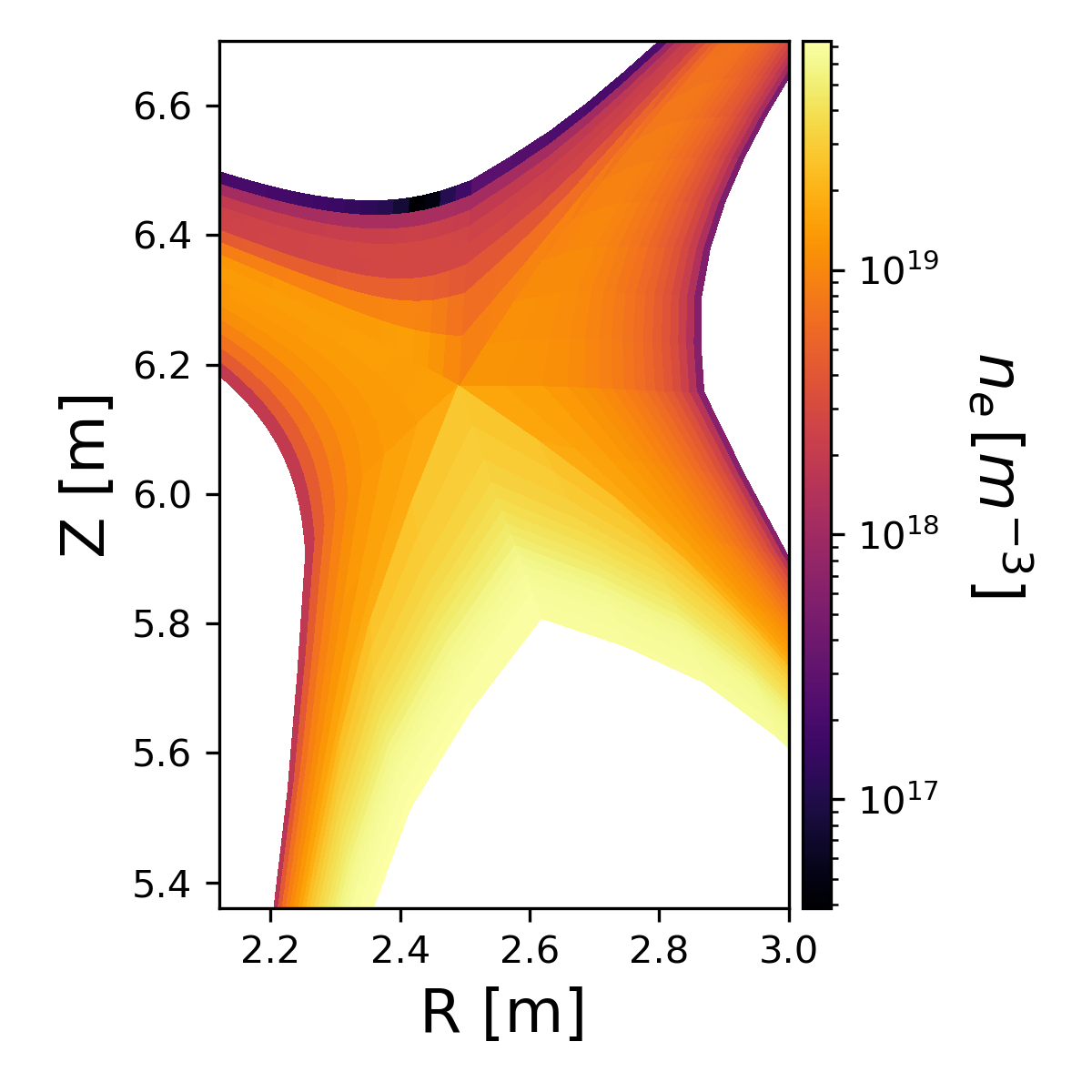}
    }\\
    \subfloat[\label{fig:step_tempzoom}]{
        \includegraphics[width=0.33\textwidth, valign=t]{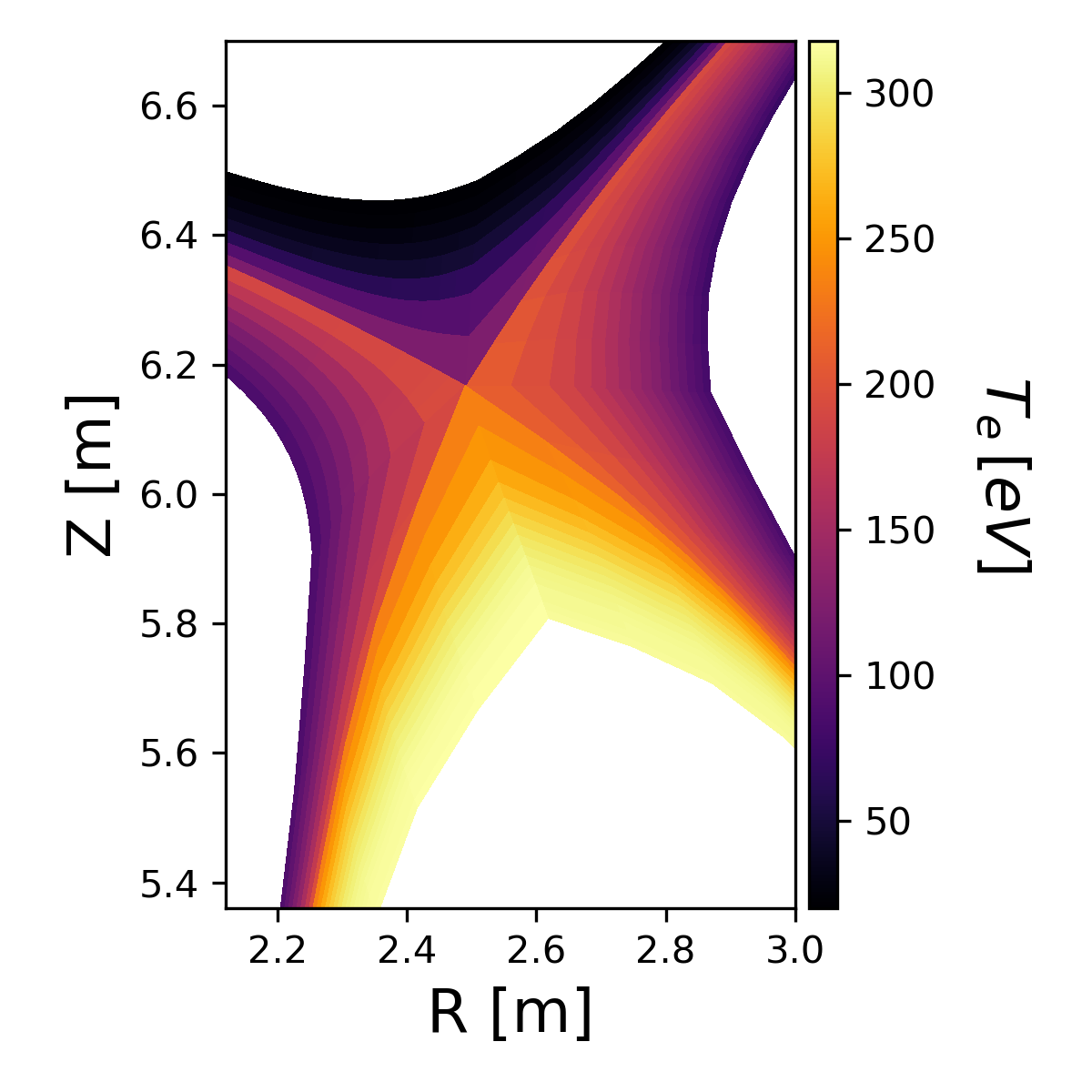}
    }
    \end{flushright}
    \vspace{1cm}
    \caption{ Simulation results from a 2D, axisymmetric simulation of the Spherical Tokamak for Energy Production. The poloidal projection of the electron density and temperature are shown in (a) and (b) respectively. A close-up of the electron density is shown in (c) and a close-up of the electron temperature is shown in (d).
    \label{fig:step_moments} }
\end{figure}

To demonstrate the effectiveness of our algorithm we conduct a 2-dimensional axisymmetric simulation in the magnetic geometry of STEP with the grid shown in Fig.~\ref{fig:stepgrid}. The simulation consists of a deuterium plasma with 100MW of input power and a particle input of $1.3\times 10^{24}$~m$^{-3}$s$^{-1}$ . The particle and heat source is Maxwellian and is present only in the innermost radial cell of the core. Within this first radial cell the particle input rate and temperature of the source is uniform. As is typically done in axisymmetric divertor design codes, an ad-hoc diffusivity is chosen to mimic turbulence which is absent in 2D simulations. Here we choose a particle diffusivity of $D=0.6$~m$^2/$s and a heat diffusivity of $\chi = 0.9$~m$^2/$s to target a heat flux width of $2$~mm. The simulation setup is similar to those in~\cite{Shukla25} where more details on \gke's gyrokinetic model can be found. For simplicity, in this test case we run without magnetic drifts or the $\mathbf E\times \mathbf B$ drift. In Fig.~\ref{fig:step_moments} we show the electron density and temperature from the simulation's steady state which is reached at $t=0.87$~ms. In these figures we can see that the simulation is well-behaved near the X-point; the electron temperature and density do not diverge.

\begin{figure}
    \centering
    \subfloat[\label{fig:stepJc_full}]{
    \includegraphics[width=0.45\textwidth, valign=t]{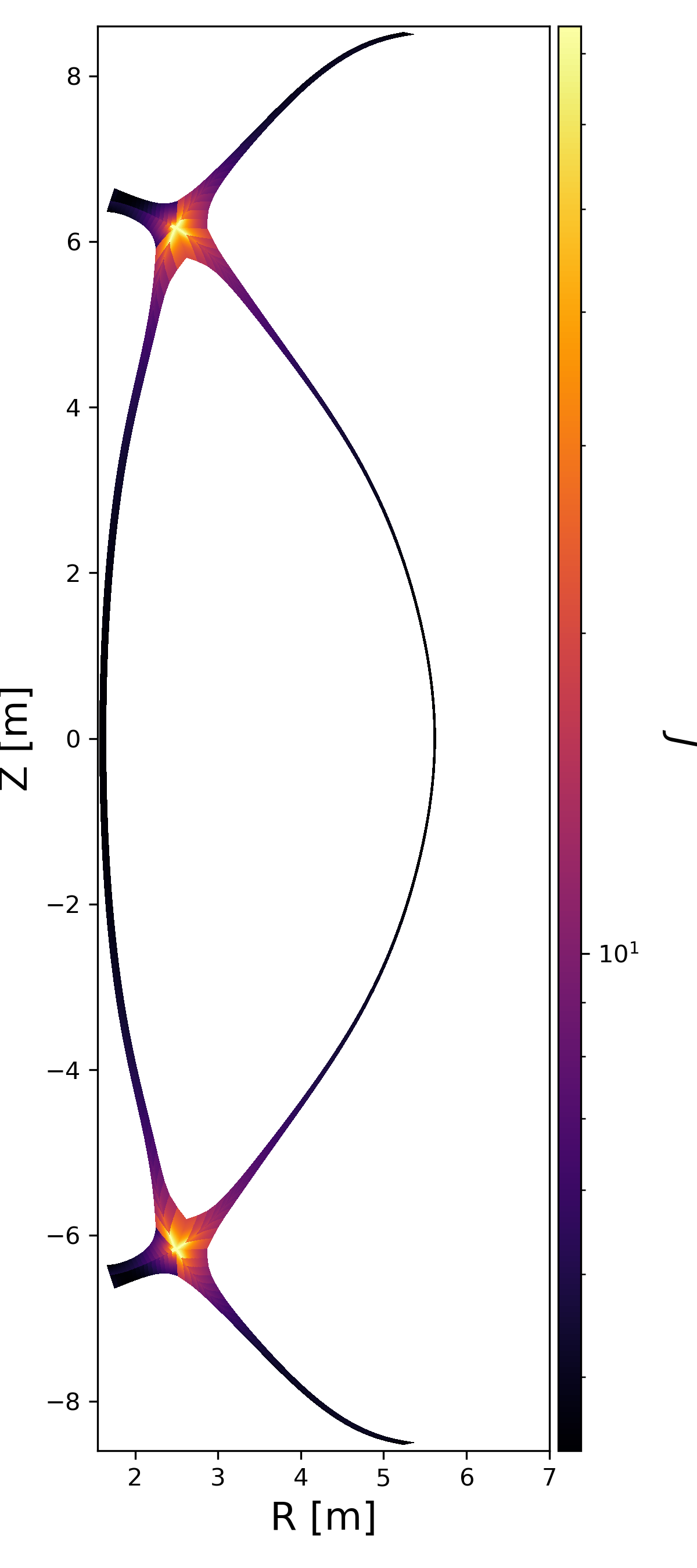}
    }
    \subfloat[\label{fig:stepJc_zoom}]{
    \includegraphics[width=0.45\textwidth, valign=t]{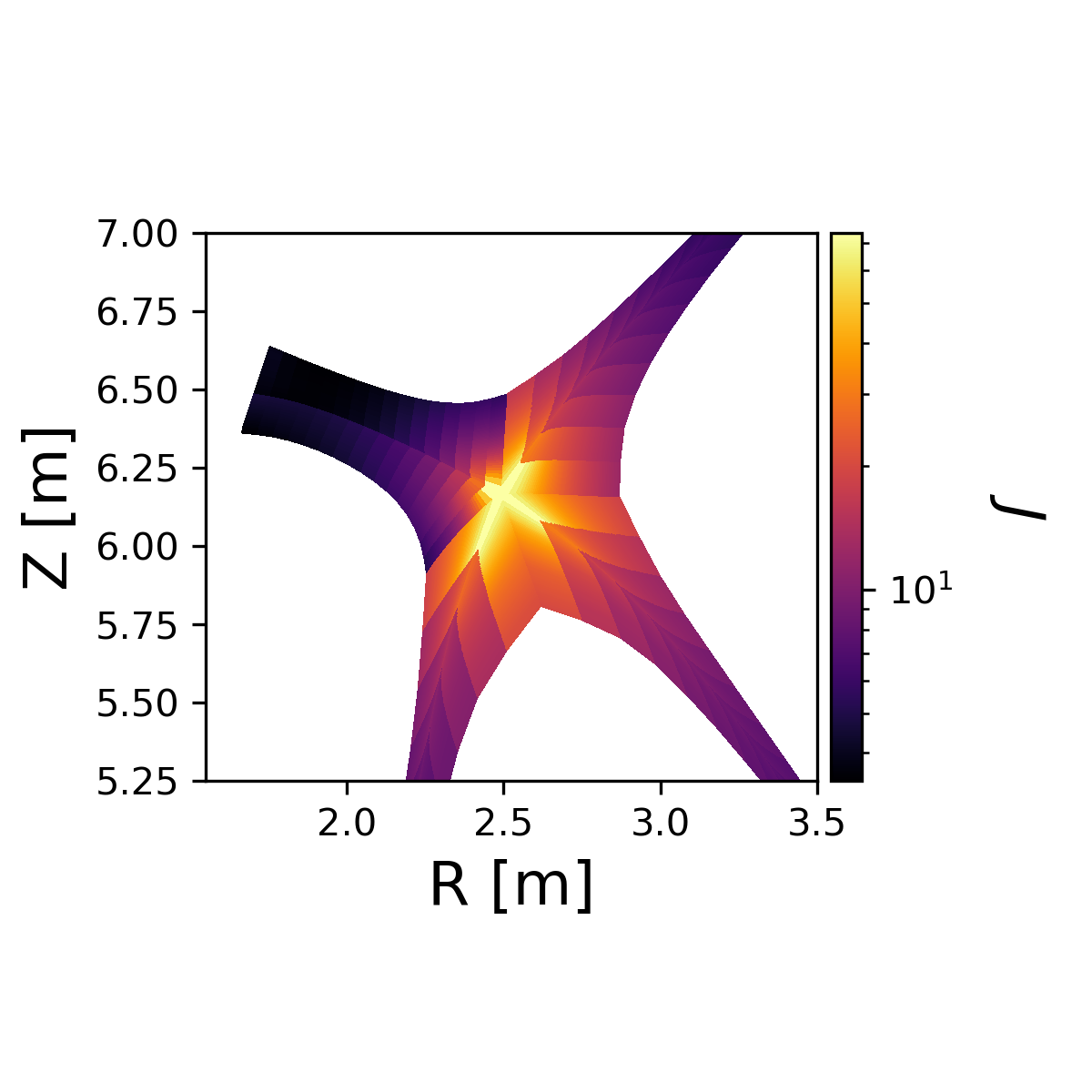}
    }
    \caption{ The poloidal projection of the Jacobian of the full domain is shown in (a) and a zoom-in around the X-point is shown in (b).
    \label{fig:stepJc} }
\end{figure}

We also plot the Jacobian, which enters into the equations for the characteristics Eq. ~\ref{eq:z1}, ~\ref{eq:z3}, and ~\ref{eq:vpar}, mapped to the poloidal plane in Fig.~\ref{fig:stepJc}. The Jacobian increases sharply approaching the X-point from all directions, but remains finite. Note that we have interpolated the values of the Jacobian to fill out the color plot in Fig.~\ref{fig:stepJc}, so it appears as if we have defined a value of the Jacobian at the X-point. However, as described in Sec.~\ref{sec:discretization} our algorithm only defines the Jacobian at interior and surface quadrature points and never directly at the X-point.

\section{Conclusion}

Field-aligned coordinate systems offer a computational advantage when conducting simulations of tokamaks because they allow for coarse resolution along the field line and larger time steps. However, using field aligned coordinates for simulations that cover both the open and closed field line regions in diverted geometries can be difficult because of the coordinate singularity at the X-point.
Here we have presented a grid generation algorithm along with a phase space discretization scheme that allows for the evolution a gyrokinetic system in X-point tokamak geometries while taking advantage of a field aligned coordinate system.

Our grid generation algorithm described in section~\ref{sec:computation} splits the domain of a tokamak into topologically distinct regions for field line tracing and then further splits the domain at the X-points resulting in a multi-block grid that ensures cell corners lie on the X-point. This grid generation algorithm uses highly accurate integrators for field line tracing and allows for direct calculation of the metric coefficients and other geometric quantities required for evolving the gyrokinetic equation in field aligned coordinates.
In section~\ref{sec:discretization} we describe the key feature of our algorithm that avoids the coordinate singularity at the X-point. Geometric quantities are evaluated at interior and surface quadrature points which do not touch the X-point and thus do not diverge.
In the final section, section~\ref{sec:example}, we demonstrate that our algorithm is capable of 2D axisymmetric simulations of X-point geometries with a \gke\ simulation of a deuterium plasma in the STEP magnetic geometry. In the future we hope to use \gke's axisymmetric solver as a complement to fluid divertor design codes and highlight the importance of kinetic effects in divertor design.

Planned improvements to our grid generation methods involve refining our grids and extending these methods for use 3D turbulence simulations. As can be seen in Fig. 4 and elsewhere, the grid spacing becomes coarse near the X-point.  There are several ways this could be improved in the future.  One is by using mesh refinement near the X-point.  Another is to use a non-uniform spacing of the $\psi$ grid to give more uniform spacing in real space near the X-point, and merge adjacent DG cells away from X-point if they become more narrow than needed (which would reduce the time step due to the Courant limit).  Another approach could be to switch to a non-aligned grid near the X-point as COGENT does.

We believe the methods described here will also work for 3D turbulence simulations; our method of evaluating geometric quantities and surface fluxes will still avoid the X-point. For 3D simulations, one would have to apply twist-and-shift boundary conditions at the parallel block boundaries~\citep{ManaTS} (for example the boundary between blocks 11 and 12 and the boundary between blocks 2 and 3 in Fig.~\ref{fig:stepgrid}). The extension to 3D will be presented in a future work. Detailed physics studies with the grids and algorithms described here, including the effect of neutrals, will also be presented in other publications.

\bibliographystyle{jpp}

\bibliography{main}

\appendix

\section{Geometric Consistency for Advection Equations}
\label{sec:appendix}

Consider the advection equation
\begin{align}
  \pfrac{f}{t} + \nabla\cdot(\mvec{v} f) = 0 \label{eq:advect}
\end{align}
where $\mvec{v}(\mvec{x},t)$ is a specified velocity profile, and
$f(\mvec{x},t)$ is a scalar advected quantity. We wish to solve this
equation on a non-rectangular domain, including on domains that can't
be covered by a single coordinate map.

For this, introduce computational coordinates $(z^1,z^2,z^3)$ and the
mapping from computational to physical space, as described in the main
text, using $\mvec{x} = \mvec{x}(z^1,z^2,z^3)$. Transforming the
advection equation to this new coordinate system we get
\begin{align}
  \pfrac{f}{t} 
  + \frac{1}{J_c}\pfraca{z^i}\left(J_c \dbasis{i}\cdot\mvec{v} f \right) 
  = 
  0. \label{eq:advect-cs}
\end{align}
In this equation now $f = f(z^i,t)$. 
\begin{figure}
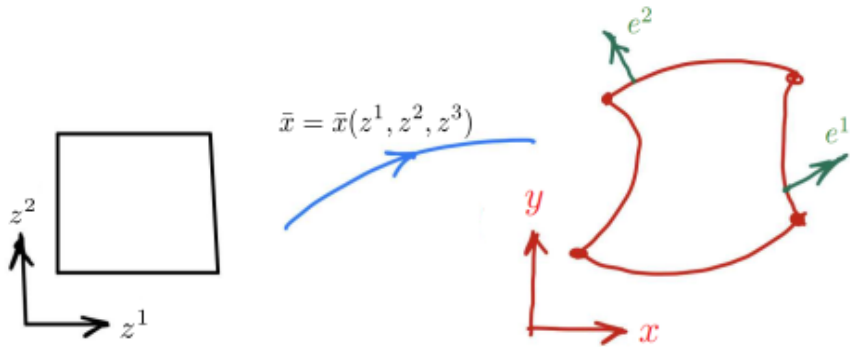

  \incfig{figures/mapc2p_convert}
  \caption{The mapping $\mvec{x} = \mvec{x}(z^1,z^2,z^3)$ maps a
    rectangle in computational space to a region in physical
    space. All information needed to compute the various geometric
    quantities in physical space (volumes, surface-areas, etc) are
    contained in the mapping function. The tangent vectors $\basis{i}$
    are tangent, and dual vectors $\dbasis{i}$ are normal, to the
    mapped faces.}
  \label{fig:mapc2p}
\end{figure}

Now consider a rectangular region in computational space. In general,
this will be mapped to a non-rectangular region in physical space. See
Fig.\thinspace(\ref{fig:mapc2p}). The volume of the mapped region is
\begin{align}
  \mathcal{V} = \iiint J_c \thinspace dz^1dz^2 dz^3.
\end{align}
In this expression we see the appearance of the ``volume element'' in
computational space $dz^1dz^2 dz^3$. This \emph{need not} have the
units of volume in physical space. However, the quantity
$J_c \thinspace dz^1dz^2 dz^3$ \emph{does} have the units of volume
(i.e length cubed).

The surface area of a face, for example, the surface that corresponds
to $z^1 = \textrm{constant}$, can be computed from
\begin{align}
  \mathcal{S}_1 
  =
  \iint \| \dbasis{1} \| J_c \thinspace dz^2 dz^3
  =
  \iint \| \basis{2}\times\basis{3} \| \thinspace dz^2 dz^3.
\end{align}
Hence, we can interpret $\| \dbasis{1} \| J_c$ or
$\| \basis{2}\times\basis{3} \|$ as the \emph{surface Jacobian} for
the $z^1 = \textrm{constant}$ face. The computational space ``surface
element'' $dz^2 dz^3$ \emph{need not} have the units of area in
physical space, but of course
$\| \dbasis{1} \| J_c \thinspace dz^2 dz^3 = \|
\basis{2}\times\basis{3} \| \thinspace dz^2 dz^3$ does (i.e length
squared)\footnote{The length of an edge of course is simply given by,
  for example, $l_1 = \int \| \basis{1} \| \thinspace dz^1$. Here, the
  quantity $ \| \basis{1} \| \thinspace dz^1$ has the units of
  length.}.

\subsection{Finite-Volume Schemes}

Before considering the discontinuous Galerkin scheme, consider a
finite-volume approach to solve \eqr{\ref{eq:advect}} directly. For
this we will assume that we have divided the domain into hexahedral cells
with flat, quadrilateral, faces. Integrate over a single such cell,
$\Omega$, to get the weak-form
\begin{wrapfigure}{L}{0.25\textwidth}
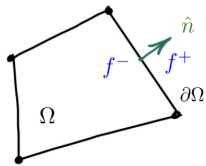

\incfig{figures/fv-cell_convert} 
\caption{A hexahedral cell $\Omega$, with surface $\partial\Omega$ and
outward-pointing surface normal $\mvec{n}$.}
\label{fig:fv-cells}
\end{wrapfigure}
\begin{align}
  \frac{d}{dt} \int_\Omega f(\mvec{x},t) \thinspace d^3\mvec{x}
  +
  \oint_{\partial \Omega} \mvec{v}\cdot\mvec{n} \hat{f} \thinspace ds
  =
  0. \label{eq:fv-scheme-xs}
\end{align}
Here, $\mvec{n}$ is the unit normal vector on the faces
$\partial\Omega$ that bound the cell. (We have six such faces and the
$\oint$ is a short-hand for computing the integrals over each
individual face). The value $\mvec{v}\cdot\mvec{n} \hat{f}$ is the
\emph{numerical flux} on the face: this is computed using the values
of $f$ on each side of the face. See
Fig\thinspace(\ref{fig:fv-cells}).

This numerical flux can be computed from\footnote{This form of the
  flux is often called the ``Lax flux''. As written we are assuming
  the $\mvec{v}$ is a continuous vector field and hence we only have a
  single value of the normal component of the velocity at the face.}
\begin{align}
  \mvec{v}\cdot\mvec{n} \hat{f}
  =
  \mvec{v}\cdot\mvec{n} \llbracket f \rrbracket
  -
  | \mvec{v}\cdot\mvec{n} | \{ f \}
\end{align}
where $\llbracket f \rrbracket = (f^+ + f^-)/2$ and
$\{ f \} = (f^+ - f^-)/2$. Note that this numerical flux is
\emph{consistent}, i.e. when $f^+ = f^- = f$ (where $f$ is the common
value) then the numerical flux is identical to the physical flux
$\mvec{v}\cdot\mvec{n} f$. This consistency condition is
\emph{critical to ensure we are actually solving the correct equations
  and getting the physics right}.

If we sum \eqr{\ref{eq:fv-scheme-xs}} over all hexahedra in the
domain, then as we have used a common flux value across each face, we
will get that the discrete finite-volume scheme conserves total
particles in the domain. Note an important point: conservation of
particles \emph{did not depend on using a consistent flux}, just a
\emph{common} flux across each face. Hence, \emph{flux consistency} is
an \emph{independent} property from conservation.

Now consider a finite-volume scheme derived from the
\emph{transformed} advection equation \eqr{\ref{eq:advect-cs}}. For
this, we multiply by $J_c$ and integrate over a single, rectangular,
\emph{computational space} cell $I_{ijk} =
[z^1_{i-1/2},z^1_{i+1/2}]\times [z^2_{j-1/2},z^2_{j+1/2}] \times [z^3_{k-1/2},z^3_{k+1/2}]$ to get
\begin{align}
  \frac{d}{dt} \iiint_{I_{ijk}} f J_c \thinspace dz^1 dz^2 dz^3
  +
  \iint_{S_{jk}}  J_c \dbasis{1}\cdot\mvec{v} \hat{f} \thinspace dz^2 dz^3
  \bigg |^{i+1/2}_{i-1/2}
  + \ldots
  =
  0. \label{eq:fv-scheme-cs}
\end{align}
Here, $S_{jk}$ is the $z^1 = \textrm{constant}$ face of cell
$I_{ijk}$. We have dropped the other surface terms from the $z^2$- and
$z^3$ directions, but they can be treated in the same way. In general,
this scheme is different than one obtained from
\eqr{\ref{eq:fv-scheme-xs}}.

However, comparing the two schemes we see a few key points. First, the
quantity $J_c dz^1 dz^2 dz^3$ that appears in the first term in
\eqr{\ref{eq:fv-scheme-cs}} is needed to compute the total number of
particles in the \emph{mapped} physical cell. Second, the $J_c$ that
appears in the surface terms must be chosen very carefully: when
combined with the dual vector into $J_c \dbasis{1} dz^2 dz^2$ it must
give the \emph{surface area} element on the $z^1 = \textrm{constant}$
surface. It must hence be the \emph{same} as used in computing the
surface area of the cell in physical space. Further, the value of
$J_c$ (and of course the $\dbasis{1}$ vector) must the same as
computed from the two cells that are attached to the $S_{jk}$
face. Hence, in a way, it is more accurate to write the surface term
as
\begin{align}
  \iint_{S_{jk}}  \hat{f} \mvec{v}\cdot \ndbasis{1}
  \thinspace ( J_c \| \dbasis{1} \| dz^2 dz^3 )
  \bigg |^{i+1/2}_{i-1/2}
\end{align}
where $\ndbasis{1}$ is the normalized dual vector which is, of course,
the unit normal to the $S_{jk}$ surface in physical space.  In this
form, the quantity $J_c \| \dbasis{1} \| dz^2 dz^3$ appears as the
purely geometric quantity with units of area. For use in the surface
term we can use Lax fluxes:
\begin{align}
  \mvec{v}\cdot \ndbasis{1} \hat{f}
  =
  \mvec{v}\cdot\ndbasis{1} \llbracket f \rrbracket
  -
  | \mvec{v}\cdot\ndbasis{1} | \{ f \}.
  \label{eq:lax-flux-cs}
\end{align}

\subsection{Discontinuous Galerkin Schemes}

To derive a DG scheme let $\mathcal{U}$ be a finite-dimensional
function space defined in each computational space cell
$I_{ijk}$. Then, a DG scheme is one for which, for all
$\psi \in \mathcal{U}$ we have
\begin{align}
  \iiint_{I_{ijk}} \psi \pfraca{t} (f J_c)_h \thinspace dz^1 dz^2 dz^3
  &+
  \iint_{S_{jk}}  \psi^{-} J_c \dbasis{1}\cdot\mvec{v} \hat{f} \thinspace dz^2 dz^3
  \bigg |^{i+1/2}_{i-1/2}
  + \ldots \notag \\
  &-
  \iiint_{I_{ijk}} \pfrac{\psi}{z^i} (\dbasis{i}\cdot\mvec{v})_h (f J_c)_h \thinspace dz^1 dz^2 dz^3
  =
  0. \label{eq:dg-scheme-cs}  
\end{align}
We have shown only a single surface term, but the other terms can be
treated identically. Note that the DG scheme determines the projection
$(f J_c)_h$ on basis functions and not $f_h$ directly
. A weak-form of the latter can be obtained by
solving the weak equality
\begin{align}
  (f J_c)_h \doteq f_h J_{ch}
\end{align}
where $J_{ch}$ is the projection of the Jacobian on the basis
functions, in terms of which the physical cell volume is given by
\begin{align}
  \mathcal{V}_{ijk} = \iiint_{I_{ijk}} J_{ch} \thinspace dz^1 dz^2 dz^3.
\end{align}
To compute the surface terms we again group it as we did in the
finite-volume case:
\begin{align}
  \iint_{S_{jk}} w^{-} \hat{f} \mvec{v}\cdot \ndbasis{1}
  \thinspace ( J_c \| \dbasis{1} \| dz^2 dz^3 )
  \bigg |^{i+1/2}_{i-1/2}
\end{align}
The numerical flux can be computed using Lax fluxes,
\eqr{\ref{eq:lax-flux-cs}}. However, in computing the surface integral
we must be careful and ensure that the
$J_c \| \dbasis{1} \| dz^2 dz^3$ that appears in it, when integrated
over the surface gives the surface area of the $S_{jk}$ face of the
$I_{ijk}$ cell in physical space:
\begin{align}
  \mathcal{S}_{jk}
  =
  \iint_{S_{jk}} J_c \| \dbasis{1} \| \thinspace dz^2 dz^3.
  \label{eq:surf-area-dg}
\end{align}

The source of aliasing error is now apparent: to compute the numerical
flux we need the \emph{value} of $f$ at a node, but we only have the
volume projection $(f J_c)_h$. One way to compute this quantity is to
evaluate $(f J_c)_h$ at the node and then divide out by the
\emph{value} of $J_c$ at that node, this being the same \emph{value}
we used in \eqr{\ref{eq:surf-area-dg}}. Note that this value is
\emph{not, and must not} be the evaluation of the volume expansion of
$J_{ch}$ at the surface node! As the expansions of the Jacobian in the
two cells that share a face may be \emph{discontinuous}, this would
lead, in general, to discontinuous surface values, and hence a
\emph{geometric inconsistent} flux across the face.


\subsection{Multi-block Geometric Consistency}

\begin{figure}
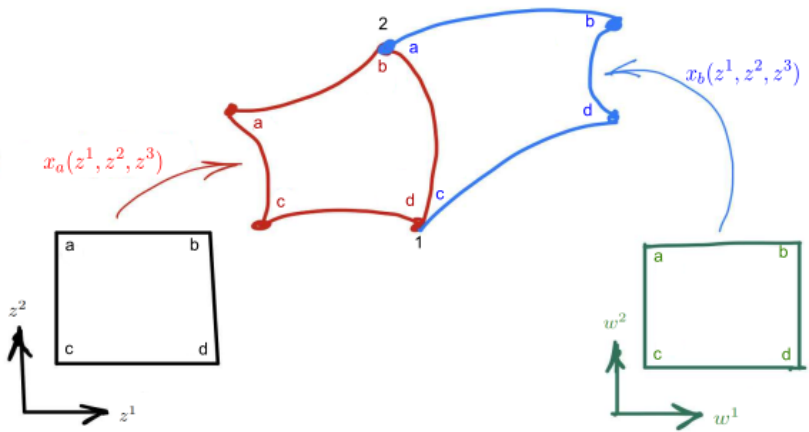

  \incfig{figures/mblock-consist_convert}
  \setkeys{Gin}{width=0.75\linewidth,keepaspectratio}
  \caption{In this two-block case we have two mappings
    $\mvec{x}_a = \mvec{x}_a(z^1,z^2,z^3)$ and
    $\mvec{x}_b = \mvec{x}_b(w^1,w^2,w^3)$. These mappings must be
    \emph{consistent} at their shared surfaces: the normals computed
    from the two maps must be parallel and the tangent vectors on the
    common surface must span the same 2D space.}
  \label{fig:mbock}
\end{figure}

For many problems on complex geometry we need to use multiple maps
of coordinate mappings (see Fig.\thinspace\ref{fig:mbock}). Let us
assume we have two mappings
\begin{subequations}
\begin{align}
  \mvec{x}_a &= \mvec{x}_a(z^1,z^2,z^3) \\
\mvec{x}_b &= \mvec{x}_b(w^1,w^2,w^3).
\end{align}
\end{subequations}
From these we can compute the tangent vectors $\basis{i}$,
$\gbasis{g}{i}$ and their duals, $\dbasis{i}$, $\gdbasis{g}{i}$ in the
usual way.
Now consider the common face marked in the figure by the line
$(1,2)$. This is a $z^1 = \textrm{constant}$ and
$w^1 = \textrm{constant}$ surface. At this surface we have some
geometric consistency conditions. Let $\basis{2}$ and $\basis{3}$ be
the tangent vectors on this surface as computed using the $\mvec{x}_a$
mapping, and $\gbasis{g}{2}$ and $\gbasis{g}{3}$ be the tangent vectors
on this surface as computed using the $\mvec{x}_b$ mapping. Then we
must ensure that
\begin{align}
  \gbasis{g}{2}\cdot(\basis{2}\times\basis{3})
  =
  \gbasis{g}{3}\cdot(\basis{2}\times\basis{3})
  =
  0.
\end{align}
This merely states that the linear spaces spanned by
$(\basis{2},\basis{3})$ and $(\gbasis{g}{2}, \gbasis{g}{3})$ at each
point on the shared surface are the same. We must also have that
$\dbasis{1}\times\gdbasis{g}{1} = 0$. Further, as the surface is
shared, the surface area computed from either mapping must be the
same:
\begin{align}
  \mathcal{S}
  =
  \iint_1^2 J_{ca} \| \dbasis{1} \| \thinspace dz^2 dz^3
  =
  \iint_1^2 J_{cb} \| \gdbasis{g}{1} \| \thinspace dw^2 dw^3
  \label{eq:surf-match-mb}
\end{align}
where $J_{ca}$ and $J_{cb}$ are the Jacobians computed from the
$\mvec{x}_a$ and $\mvec{x}_b$ mappings respectively,

The geometric consistency is easier to impose on the tangent and dual
vectors than it is on the surface areas. The latter is particularly
tricky when doing DG schemes: we must ensure that the nodal points we
choose based on the $z^i$ and $w^i$ coordinates on shared surfaces
actually map to the \emph{same physical point} in physical space. This
need not need be the case as, in general, each of the two mappings may
be nonlinear and different. Further, we must ensure that the condition
\eqr{\ref{eq:surf-match-mb}} must be satisfied
\emph{exactly}. Otherwise we will introduce an error in particle flux
across block boundaries, potentially causing long-term instability in
the scheme.

\end{document}